\documentclass[onecolumn]{aastex631}

\usepackage{graphics}
\usepackage{grffile}
\usepackage{rotating}

\usepackage{natbib}
\usepackage{url}

\usepackage{amsmath}  
\usepackage{amssymb,amsmath}  
\usepackage{color}
\usepackage{times}
\usepackage[normalem]{ulem}             

\usepackage{cancel}
\usepackage{verbatim}                           
\usepackage{mathrsfs}                           
\usepackage[mathscr]{euscript}
\usepackage{fancyref}
\usepackage{hyperref}
\usepackage{physics}

\usepackage{ira_definitions}

\usepackage{jim_definitions}

 \newcommand{\FTA}{\widetilde A(f)}
 \newcommand{\amean}{\langle a \rangle}
\newcommand{\abmean}{\langle a_b \rangle}

\newcommand{\pFP}{p_{\rm fp}}
\newcommand{\Smin}{S_{\rm min}}
\newcommand{\Smean}{\langle S(f)\rangle}
\newcommand{\cycfunc}{Q}

\newcommand{\RoneTNS}{FRB~20121102A}
\newcommand{\RthreeTNS}{20180916B}

\begin{document}

\title{
Empirical Assessment of Aperiodic and Periodic  Radio Bursts from Young Precessing Magnetars} 

\shorttitle{Spin and Precession}

\author[0000-0002-4049-1882]{J. M. Cordes}
\affil{Cornell Center for Astrophysics and Planetary Science and Department of Astronomy, Cornell University, Ithaca, NY 14853, USA}
\author[0000-0001-6618-6096]{I. Wasserman} 
\affil{Cornell Center for Astrophysics and Planetary Science, Department of Astronomy, Department of Physics,  Cornell University, Ithaca, NY 14853, USA}
\author[0000-0002-2878-1502]{Shami Chatterjee}
\affil{Cornell Center for Astrophysics and Planetary Science and Department of Astronomy, Cornell University, Ithaca, NY 14853, USA}
\author[0000-0002-3208-665X]{G. Batra} 
\affil{Cornell Center for Astrophysics and Planetary Science and Department of Physics, Cornell University, Ithaca, NY 14853, USA}

\correspondingauthor{J. M. Cordes}
\email{jmc33@cornell.edu}

\begin{abstract}
We analyze the slow periodicities identified  in burst sequences from \Rone\ and \Rthree\ with periods of about 16 and 160~d, respectively, while also addressing the absence of any fast periodicity that might be associated with the spin of an underlying  compact object. 
 Both phenomena can be accounted for by a young, highly magnetized, precessing neutron star  that  emits beamed radiation  with significant imposed phase jitter.  Sporadic narrow-beam emission  into an overall  wide solid angle can account for the necessary phase jitter, but the slow periodicities with 25 to 55\% duty cycles constrain beam traversals  to be significantly  smaller.  Instead,  phase jitter  may result from  variable emission altitudes that  yield large retardation and aberration delays.   A detailed arrival-time analysis for triaxial precession   includes wobble of the radio beam and the likely larger,  cyclical torque resulting from the changes in the spin-magnetic moment angle.   These effects will  confound identification of the fast periodicity 
 in sparse data sets  longer than about a quarter of  a precession cycle unless fitted for and removed  as with orbital fitting.   Stochastic spin noise, likely to be much larger than in radio pulsars, may  hinder detection of any fast-periodicity in data spans longer than a few days.  These decoherence effects will dissipate as  FRB sources age, so they may evolve into objects with properties similar to Galactic magnetars. 
\end{abstract}


\section{Introduction}
\label{sec:intro}

Fast radio bursts (FRBs) are well established as an extragalactic phenomenon (for reviews see
\citet[][]{2019A&ARv..27....4P,2019ARA&A..57..417C}) that now has also been seen in a Galactic magnetar
\citep[][]{2020Natur.587...59B}. Precise interferometric localizations of some of these bursts have allowed the identification of host galaxies and the measurement of their redshifts. The most distant FRB sources imply  enormous radio burst energies, requiring not only a coherent emission process, but also an efficient one that causes radiation reaction on the emitting particles.  

Some sources have  repeated episodically while others  have produced only a single detected burst,
so far.
The first and best studied repeating source \RoneTNS\  (hereafter R1 or \Rone; \citealt{ssh+16a}) is located in  a dwarf, star-forming  galaxy at a redshift $z=0.193$ \citep{clw+17, tbc+17}. It has shown  highly intermittent bursts, occasionally appearing at a rate of tens per hour, but more typically showing only a few events in a daily observing session interleaved with long quiet periods (days to months) \citep[][]{2016ApJ...833..177S,2017ApJ...850...76L}. Even on  days when multiple bursts are seen with sub-second spacings, no fast periodicity has been identified  \citep[][]{clw+17,2018ApJ...863....2G,2018ApJ...866..149Z,2021Natur.598...267L}.
Nor has 
the repeater \RthreeTNS\ \citep[hereafter R3 or \Rthree;][]{2019ApJ...885L..24C} shown a fast periodicity (2~ms to $\sim$seconds) in burst sequences \citep[e.g.][]{2020ApJ...896L..41C}. R3 has also been localized by \citet{2020Natur.577..190M} to a star-forming region in a massive spiral galaxy at a redshift $z = 0.034$.
Very recently, a quasi-periodicity with a
 0.22~s period has been reported for a short burst sequence from FRB~20191221A
\citep[][]{2021arXiv210708463T}. 
It remains to be seen whether the FRB will show any long-term periodicity.

For R1 and R3, however, long-term periodicities  have been identified.  
R3 showed a slow  periodicity  in a train of 38 detected bursts \citep[][]{2020Natur.582..351C}
in the form of a quasi-periodic $\sim$5-day window repeating every 16.4 days within which detected bursts occur, though there are windows in which no bursts are detected.  The initial uncertainty about whether the apparent 16.4~day period was a multiple of the true period has been removed by additional observations  \citep[][]{2020MNRAS.499L..16M}. 

A similar slow periodicity  subsequently was  identified from R1 \citep[][]{2020MNRAS.495.3551R}  with a period of 157$\pm$7~days and a 56\% duty cycle, corroborated by  \citet[][]{2020MNRAS.tmp.3001C} who find a  161$\pm$5~day period and  a 54\% duty cycle.
A large sample of $\sim 1600$ bursts from R1 
\citep[][]{2021Natur.598...267L}
over a 60~d period is also consistent.   For other FRB sources, insufficient numbers of bursts have been reported  to assess whether such periodic detectability windows are a generic feature of repeating FRBs or not.

A wide variety of explanations for the slow periodicities has been offered \citep[e.g.][]{2021MNRAS.502.4664K}.
Their long periods are commensurate with those expected from orbital motion, such as 
a neutron star (NS) orbiting a companion  star \citep[][]{2020ApJ...893L..39L, 2020ApJ...893L..26I} or a binary that also includes  an asteroid belt around the companion \citep[][]{2020ApJ...895L...1D}.   However \citet[][]{2020MNRAS.496.3390B} associate the slow periodicity  with a very slowly spinning NS while others invoke spin precession, including 
geodetic precession \citep[][]{2020ApJ...893L..31Y}, 
forced precession \citep[][]{2020MNRAS.497.1001S}, 
and free precession \citep[][]{ 2020ApJ...892L..15Z, 2020ApJ...895L..30L}.

In this paper we analyze free precession which, 
 unlike geodetic and forced precession, causes the torque to vary over a precession cycle due to the changing angle between the spin axis and magnetic dipole axis.  This can enhance arrival time variations by a large factor over those expected from mere wobble of an emission beam. 
We  aim to explain both the absence of a fast periodicity in any of the repeating objects, especially in the large burst samples of R1,  and the presence of the slow quasi-periodicities in bursts from  R1 and R3.
The common features of R1 and R3 suggest that the absence of a fast periodicity is the norm in these kinds of objects, perhaps 
indicating that burst emission does not directly
involve the  spin of an underlying object.  This is the case  in  magnetar shock models
\citep[e.g.,][]{2017apj...843l..26b, 2017apj...841...14m, 2017apj...842...34w, 2018MNRAS.481.2407M, 2018ApJ...868L...4M, 2019MNRAS.485.4091M}, where bursts are produced from coherent emission in synchrotron masers far outside  the magnetosphere of a neutron star.  However, an alternative is that  bursts originating from the magnetosphere may have times of arrival (TOAs) where periodic emission is masked by other effects, as considered in this paper \citep[see also][]{2017MNRAS.467L..96K}.

We assume the central engine produces radio bursts  from the magnetosphere of a young magnetar as  magnetic-driven, coherent radio emission. That possibility has become more plausible \citep[][]{2020MNRAS.tmp.2850K} with the recent discovery of an extremely bright millisecond-duration radio burst from a Galactic magnetar, SGR~1935+2154 by the  CHIME telescope at  600~MHz  \citep[][]{2020ATel13681....1S, 2020arXiv200510324T} and STARE2 at 1.4~GHz \citep[][]{2020Natur.587...59B}.  
 If the Galactic magnetar was instead at the extragalactic distance of R3, the detected burst at 1.4~GHz would be similar to some of the fainter R3 bursts, strongly suggesting that magnetars are associated with at least some subset of FRBs.

If the central engine for (at least some) FRBs is associated with a young magnetar, a dense nebula will prohibit burst propagation unless supernova ejecta have expanded sufficiently \citep[][]{pir16} or if there are evacuated propagation channels produced by repeated flares from the magnetar or by  nonlinear propagation of strong electromagnetic waves \citep[e.g.][]{2020MNRAS.496.3308L}.    The spin period at the time of radio `break out' may be significantly longer than its period in the immediate aftermath of the supernova explosion and matter fallback.  The underlying spin periods of FRB sources may then be of order seconds or longer, an important consideration in free-precession models where the figure of the star determines the ratio of the precession and spin periods. 

A companion paper  \citep[][henceforth Paper I]{2022ApJ...928...53W} presents a more general and more detailed description of the requirements for having magnetic driven triaxiality of a NS and its observable consequences.     

The paper is organized as follows.   
Section~\ref{sec:spinstate}  discusses the basic picture of a highly magnetized NS with s complex surface field.
Section~\ref{sec:decohere}   discusses the conditions needed to hide fast periodicities in burst sequences.  
Section~\ref{sec:triaxial} summarizes from Paper I  the features of a precessing triaxial star needed for our discussion of slow and fast periodicities. 
Section~\ref{sec:bpmf} derives the beam precession modulation function for beamed radiation from a precessing NS. 
Section~\ref{sec:timing} presents the cyclical arrival time variations expected from the combination of a precessing beam and the variable torque acting on the NS. 
Section~\ref{sec:applications} considers the detectability of spin periodicities for precessing NS under representative condiditions.
Section~\ref{sec:conclusions} gives the summary and conclusions.
Appendix~\ref{app:spinnoise} details spindown from magnetic torques and our extrapolation of stochastic spin noise from pulsars to young magnetars.
Appendix~\ref{app:spectrum} gives a derivation for the power spectrum of bursts with phase jitter and nulling. 
Appendix~\ref{app:cyc_torque} derives the arrival-time variation from the cyclical component of the torque acting on a precessing neutron star. 

\section{Spin state of a magnetically active neutron star}
\label{sec:spinstate}

A highly magnetized object with rapid spin will most likely show large variations in spin rate and direction as the figure of the star changes.   Multiple bursts spread broadly in pulse phase may result from multiple radio beams tied to  a highly non-dipolar surface magnetic field.  A dynamic magnetic field topology will also cause changes in the directions and possibly the number of these beams as the changing figure of the star induces precession  and torque variations. 
   
\begin{figure}[t] 
   \centering
   \includegraphics[width=0.6\figwidth]{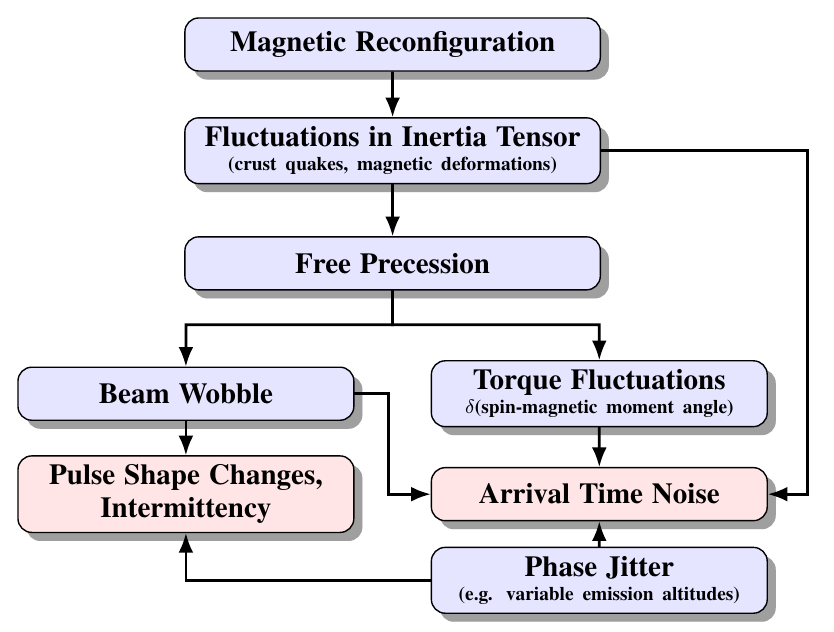}
   \caption{Diagram showing the effects from 
   stochastic  reconfiguration of a magnetar's magnetic field. [Adaptation of \citet[][Figure 4]{1993ASPC...36...43C}]. The red-shaded boxes represent observable consequences while blue shading indicates the underlying physical processes.
   }
   \label{fig:block_diagram}
\end{figure}

Figure~\ref{fig:block_diagram} diagrams the consequences of stochastic changes in magnetic field for variations in the spin rate and direction and the resulting   changes in pulse shape and departures of arrival times from strict periodicity. 
This picture raises additional questions:
Does an FRB source evolve from aperiodic to periodic emission?  In particular, do FRB sources evolve into Galactic type magnetars that episodically show periodic radio emission?   If so,  what features of the emission, if any,  are in common throughout the magnetar's lifetime?

We consider a young magnetar with a large, highly dynamic magnetic field and star figure.
To avoid over-rapid spindown to uninteresting periods, i.e. those for which crust quakes and flares are infrequent,
we assume the surface field is dominated by multipolar components well in excess of the  dipolar field. 

Isolated neutron stars spin down according to a torque law similar to that of magnetic dipole radiation.   For a characteristic spindown time
$\taus = \nu / (n-1)\dot\nu$, where 
$n\sim 2$~to 3 is the braking index {$n\sim 2$~to 3 is the braking index (and possibly as small
as 1.4 for the Vela pulsar \citep[][]{1996Natur.381..497L}, athough a more recent analysis yields 2.8 
\citep[][]{2017MNRAS.469.4183A},
the spin frequency decreases  as
$\nu(t) = \nu_0 (1 + t / {\taus}_0)^{-1/(n-1)} $ where ${\taus}_0$ is the initial spindown time at $t=0$.
Departures from $n=3$ can be for a variety of reasons; proposals include magnetic field decay 
\citep[][]{2021MNRAS.tmp.2446J}
or growth that, respectively, yield $n>3$ and $n<3$; magnetic alignment 
\citep[][]{gol70,2001aanda...376..543T} 
or counteralignment; crust-superfluid interactions, as in the Vela pulsar; and magnetospheric evolution \citep[][]{1997MNRAS.288.1049M}.       Given that FRBs may involve very young NS that do not have an internal superfluid (see Paper~I for discussion), none of these effects may apply. 
For example, \citet[][]{2021MNRAS.tmp.2446J} discuss field decay for magnetars and find a best estimate of
4~kyr for the decay time scale.   Magnetospheric evolution is a possibility but is likely to be more complex than in radio pulsars. 

Typical (`canonical') radio pulsars with $10^{12}$~G dipole fields (expressed at the NS surface with radius $R$) and $\Pspin\sim 1$~s periods have current spindown times $\taus \sim 10$~Myr, while the youngest Galactic pulsars are $\sim$kyr in age.  FRB sources may be significantly younger than this. 

Figure~\ref{fig:nu_vs_t} shows spindown curves for five cases with different birth spin frequencies and dipolar fields.   We have used a magnetic dipole spindown law ($n=3$) and a moment of inertia
$\sim 10^{45}$~g~cm$^2$, as discussed in Appendix~\ref{app:spinnoise}.  The figure demonstrates the much more rapid spindown of magnetars compared with either canonical pulsars and millisecond pulsars.     Magnetar models suggest  $t \lesssim 100$~yr for the source of FRB~121102 \citep[e.g.][]{2017apj...843l..26b}, which is known to have been  intermittently active since late 2012 \citep[][]{sch+14}, implying a minimum age of 9~yr. This age range is indicated in the figure as the shaded rectangle.     The spindown curves show that the spin frequency has remained constant for the age range of FRB~121102 if the dipolar field $\lesssim 10^{15}$~G and for birth spin rates
$\nu_0\lesssim 0.1$~Hz.  However for large fields a  rapid birth spin rate of 10~Hz declines to $\sim 1$~Hz for $B_{\rm d} = 10^{15}$~G and to $10^{-2}$~Hz for $10^{17}$~G.

\begin{figure}[b] 
   \centering
   \includegraphics[width=0.6\figwidth]{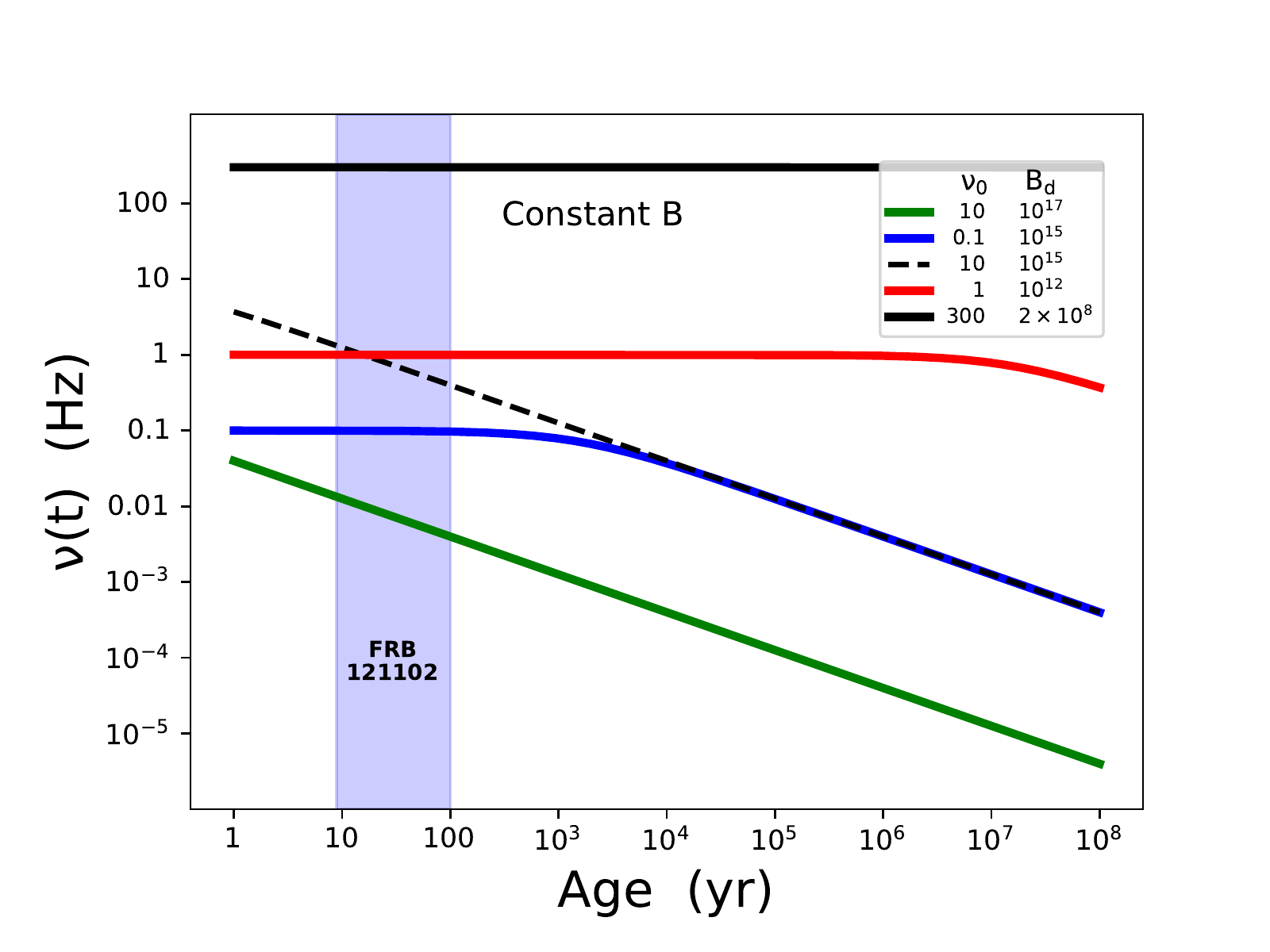}
   \caption{Spin frequency vs age for  different surface dipolar field strengths $B_{\rm d}$ (Gauss) and 
   initial spin frequencies $\nu_0$ (Hz).
   The curves evaluate the expression in the text for a braking index $n=3$.   The shaded band indicates plausible constraints on the apparent age of the FRB~121101 source ($\gtrsim 9$ to $\sim 100$~yr),  where the lower bound simply follows from the fact that the source was discovered in data acquired in 2012.}
   \label{fig:nu_vs_t}
\end{figure}

Spindown therefore cannot yield a NS with  both a large {\em dipolar} magnetic field $\gtrsim 10^{15}$~G and a spin rate faster than $\sim 1$~Hz in the age range for FRB~121102.   However it is possible to have 
a {\em total} surface field strength well in excess of $10^{15}$~G if the dipolar component makes a minority contribution at $r = R$ compared to higher-order multipoles, which fall off more rapidly with radius, contributing negligibly to the torque that is applied at the light cylinder radius, $\rlc = c / 2\pi\nu$.  

For example, the dipole and quadrupole components scale as $B_{\rm d} \propto r^{-3}$ and $B_{\rm q} \propto r^{-4}$, respectively, so the  ratio of these fields at the NS surface is
$B_{\rm q}(R) / B_{\rm d}(R) = c\sqrt{f_{\rm q}}/2\pi\nu R \simeq  (50 / \nu R_6) \sqrt{f_{\rm q} / 10^{-4}}$  where $f_{\rm q}$ is the fraction of the torque contributed by the quadrupolar field at $r = \rlc$ and the NS radius is $R = 10^6 R_6\, {\rm cm} $. 
The quadrupolar surface field  could  be as large as $ 10^{17}$~G for a surface dipole field $B_{\rm d}(R) = 10^{15}$~G and higher-order multipoles  even larger and still contribute a fraction $f_{\rm q} < 0.1$\% to the torque.   The surface field strengths we consider in this paper are well below this maximal value, so it is possible for internal and surface fields to dramatically affect the figure of the NS while not causing spindown that is too rapid. 

Figure~\ref{fig:nu_vs_t} applies to objects with constant dipolar magnetic fields.  We note that while magnetic field decay is likely not relevant for very young magnetars ($\ll 1$~kyr), older objects may have fields significantly smaller than their birth fields \citep[e.g.][]{2013MNRAS.434..123V}.   Field decay implies that spin rates will be larger than those shown in Figure~\ref{fig:nu_vs_t} for ages $\gtrsim 1$~kyr and may thus alter some of the above discussion.

\section{Decoherence of Spin-related Periodicities in Burst Sequences}
\label{sec:decohere}

Overall, we seek a model that accounts for the absence of a fast periodicity in burst sequences but allows a slow periodicity to be manifested.     
A  slow, coherent  periodicity like those observed requires  the figure of the star to remain constant and the instantaneous spin vector to be misaligned from the principal moment of inertia.   However,  stochastic changes in the star's figure will make the precession  less than perfectly coherent. 
We will first look at the requirements for masking a rapid spin period in a sequence of bursts obtained over a few hours,  a time scale for which long term precession can be ignored. 

The detectability of periodicities 
is tied to the nature of the bursts themselves: do they result from the rotational sweep of a narrow beam across the line of sight or are they infrequent temporal flashes with millisecond durations seen when the (potentially wide) beam is directed toward an observer?  
If observed burst durations measure the time for a spinning and precessing beam to cross the line of sight, the beam luminosity can persist for a longer time but no longer than about one spin period; otherwise a fast periodicity could be established from the multiple sequential bursts that would occur. 
Very fast precession of a very narrow beam could alter this conclusion, but in this paper we largely consider long precession periods  corresponding to the weeks to months of the quasi-periods discussed in the introduction.  We therefore incorporate the fact that the beam luminosity must vary 
on time scales less than and perhaps much less than the spin period.    This is the same picture as for pulsars that show strong pulse to pulse variability, including giant pulses from the Crab pulsar \cite[][]{1995ApJ...453..433L}.

The intensity is modeled as   a train of  $\Nb$ spin periods,
\be
I(\phi) =  \sum_{j=0}^{\Nb-1}   a_{j} A(\phi - \phi_{j}),
\label{eq:sigmod1}
\ee
with  bursts having identical shapes $A(\phi)$ and widths $W_{\rm A}$  but different amplitudes $a_j$, as discussed below.  The phase offset for the $j^{\rm th}$ burst is  
\be
\phi_j \equiv \phi(t_j) = \phim(t_j) + \phij(t_j) +  \phisn(t_j)  +  \phin(t_j).
\label{eq:phimodtotal}
\ee 
The first term, $\phim$, is
 deterministic and modellable, including   spindown, orbital,  and precession terms, 
\be
\phim(t) =  \phispin(t) + \phiorb(t) + \phip(t).
\label{eq:phimodfit}
\ee
Since orbit determination with pulsar timing is a well-solved problem, we exclude it from our remaining discussion.  However,  FRBs might in fact involve orbital motion that contributes to the complexity of burst sequences, especially if in conjunction with precession. The last three terms  in Eq.~\ref{eq:phimodtotal} correspond to    
stochastic phase jitter $\phij$, spin noise $\phisn$,   and   measurement  noise $\phin$.  Spin noise has nonstationary statistics, like those of a random walk \citep[e.g.][]{2010ApJ...725.1607S}, and is generated (e.g.) by stochastic changes in moment of inertia or in the magnetic torque.   By contrast, phase jitter is a random process with stationary statistics that represents variable emission phases relative to a fixed phase tied to the spinning star (or its magnetosphere) \citep[][]{1985ApJS...59..343C}.

\subsection{Dephasing of Burst Periodicities in Short Data Spans (hours)}

Highly episodic  burst sequences from \Rone\ have yielded tens to hundreds of bursts over a few hours  at some epochs \citep[][]{2018ApJ...866..149Z,  2019ApJ...876L..23H,2021Natur.598...267L}, including bursts separated by just a few seconds,  and yet no fast periodicity has been identified.   This has led  to characterization of burst occurrences as modified Poisson processes, such as the Weibull process 
\citep[e.g.][]{2018MNRAS.475.5109O, 2020MNRAS.tmp.3001C}.
However the lack  of periodicity detections does not necessarily imply  there is no underlying periodicity.   

We demonstrate that the power spectrum of a burst sequence  can be devoid of spectral lines even in the simplest case where bursts involve a strict periodicity modified only by phase jitter and where there is no additive radiometer noise.

 \subsubsection{Power Spectrum of a Continuous Burst Sequence}
 \label{sec:spec}

We assume that  burst phases $\phi_j  = j + \phi_J$ are integers augmented by  phase jitter $\phij$ that is uncorrelated between bursts but  has identical 
rms $\sigma_{\rm J} = \langle \phij^2\rangle^{1/2}$.  
We also assign  the same mean and rms for  burst amplitudes, 
$\langle a_{j} \rangle = \langle a \rangle$ 
and $\sigma_{a_j} \equiv \langle a \rangle \ma$,
which defines the modulation index $\ma$ 
 (rms amplitude divided by the mean), and angular brackets denote ensemble average. 
 
The paucity of bursts from FRB~121102 even on days when it is active
\citep[][]{ssh+16a, 2016ApJ...833..177S,2017ApJ...850...76L, 2018ApJ...863....2G,2018ApJ...866..149Z,  2019ApJ...876L..23H, 2020MNRAS.495.3551R,2020MNRAS.496.4565C,2021Natur.598...267L}
suggests that most  are too weak to detect or they have zero amplitudes (nulls).   

We therefore define a burst fraction $\fb \le 1$ as the subset of the $\Nb$ spin periods in which there is a non-zero burst amplitude. Later we also make use of the null fraction,  $\fnull = 1-\fb$.  
As defined in Appendix~\ref{app:spectrum},
the mean of all amplitudes (nulls and bursts) becomes $\amean = \fb\abmean$ where $\abmean$ is the mean 
of bursts (i.e. excluding nulls) and the modulation indices are related as $1+\ma^2 = (1+\mb^2)/\fb$.    For small $\fb\ll 1$, we have $\amean \ll \abmean$ but $\ma \gg \mb$. 
The mean number of potentially detectable bursts is  $\fb\Nb$. 

 In Appendix~\ref{app:spectrum} we derive the ensemble average spectrum of  $I(\phi)$
 with frequencies  $f$ expressed in cycles per unit spin phase, 
 \be
\Smean  &\equiv&  \langle\vert \widetilde I(f) \vert^2   \rangle 
	= 
	\sigma_a^2  \Nb \vert \FTA \vert^2
	\Bigl[
		 1 +  R_{\rm L}(f) \sum_{\ell=0}^{\infty} \Delta_{\rm \Nb}(f - \ell)   
	\Bigr], 
\label{eq:spec}
\ee 
where $\sigma_a^2 = \fb (1+\mb^2 -\fb) \abmean^2$ and the tilde represents Fourier transform. The line to continuum ratio is 
 \be
 R_{\rm L}(f) 
=  \frac{\Nb  \Jitt^2(f)}{\ma^2} 
= \frac{ \fb \Nb  \Jitt^2(f)} { 1 + \mb^2 - \fb} 
 \label{eq:lineratio}
 \ee
 and the  jitter `form factor' $\Jitt(f) = \exp[-2(\pi f \rmsj)^2]$ for a zero-mean Gaussian distribution
 (Eq.~\ref{appeq:Jitt}). 
Spectral lines are centered on integer frequencies $f = \ell$ 
with shapes  $\Delta_{\rm \Nb}(f) \le 1$ equal to a `sinc' squared function (c.f. Eq.~\ref{eq:appsinc2}).
The line width is $\Delta f \simeq \Nb^{-1} \ll 1$ for large $\Nb$.
The  continuum component and  spectral lines are shaped by 
the square of the Fourier transform of the burst shape
 $\vert \tilde A(f) \vert^2 $, which extends up to frequencies $\sim 1/W_{\rm A}$.
 The number of spectral lines comparable in amplitude to the largest at $f=1$ is $\sim 1/W_{\rm A}$ unless
they are reduced significantly by the  jitter form factor.   For large  $\sigma_{\rm J}$,  the form factor becomes very small. 

\begin{figure}[t] 
   \centering  
   \includegraphics[width=0.45\figwidth]{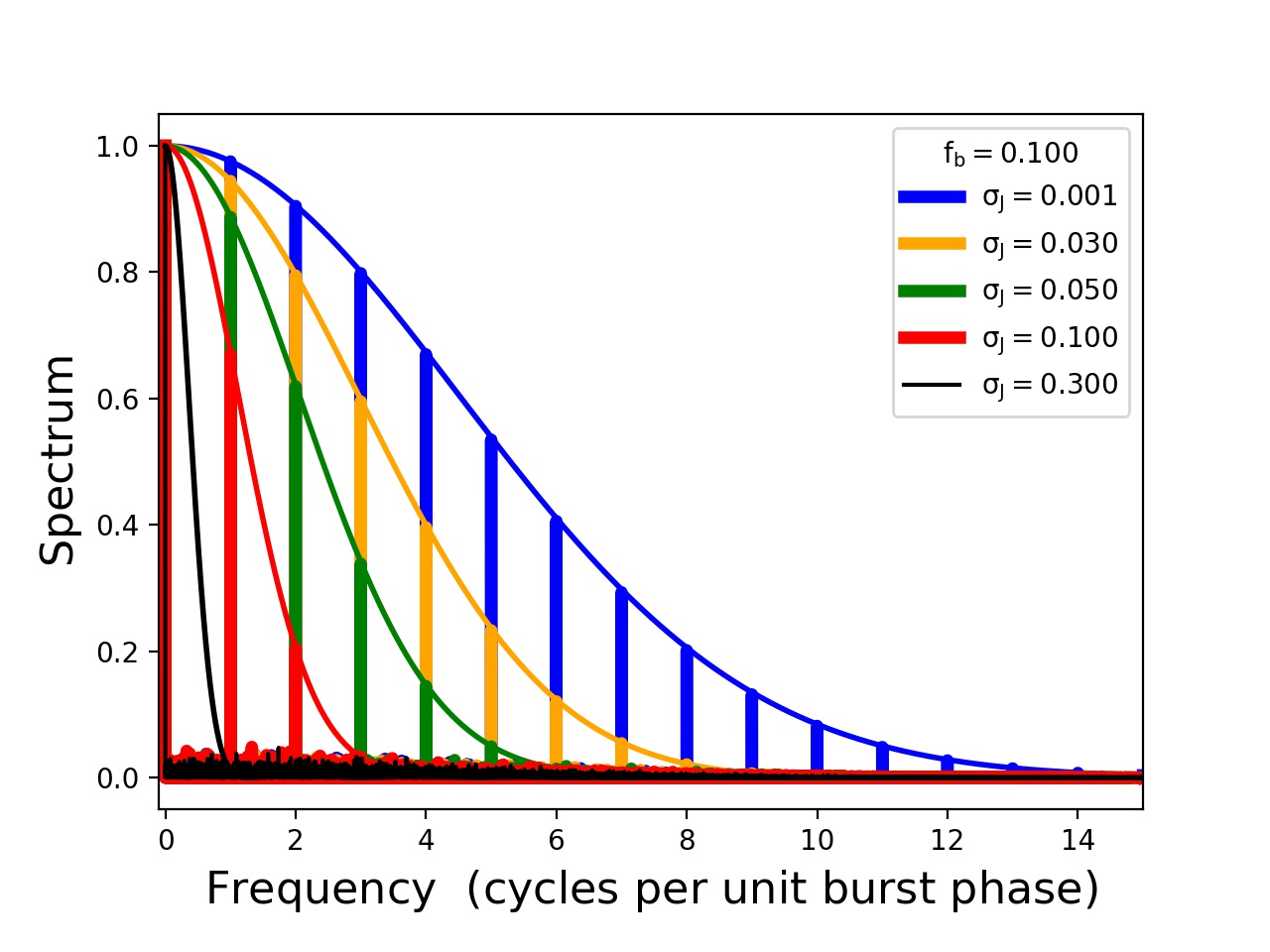} 
   \includegraphics[width=0.45\figwidth]{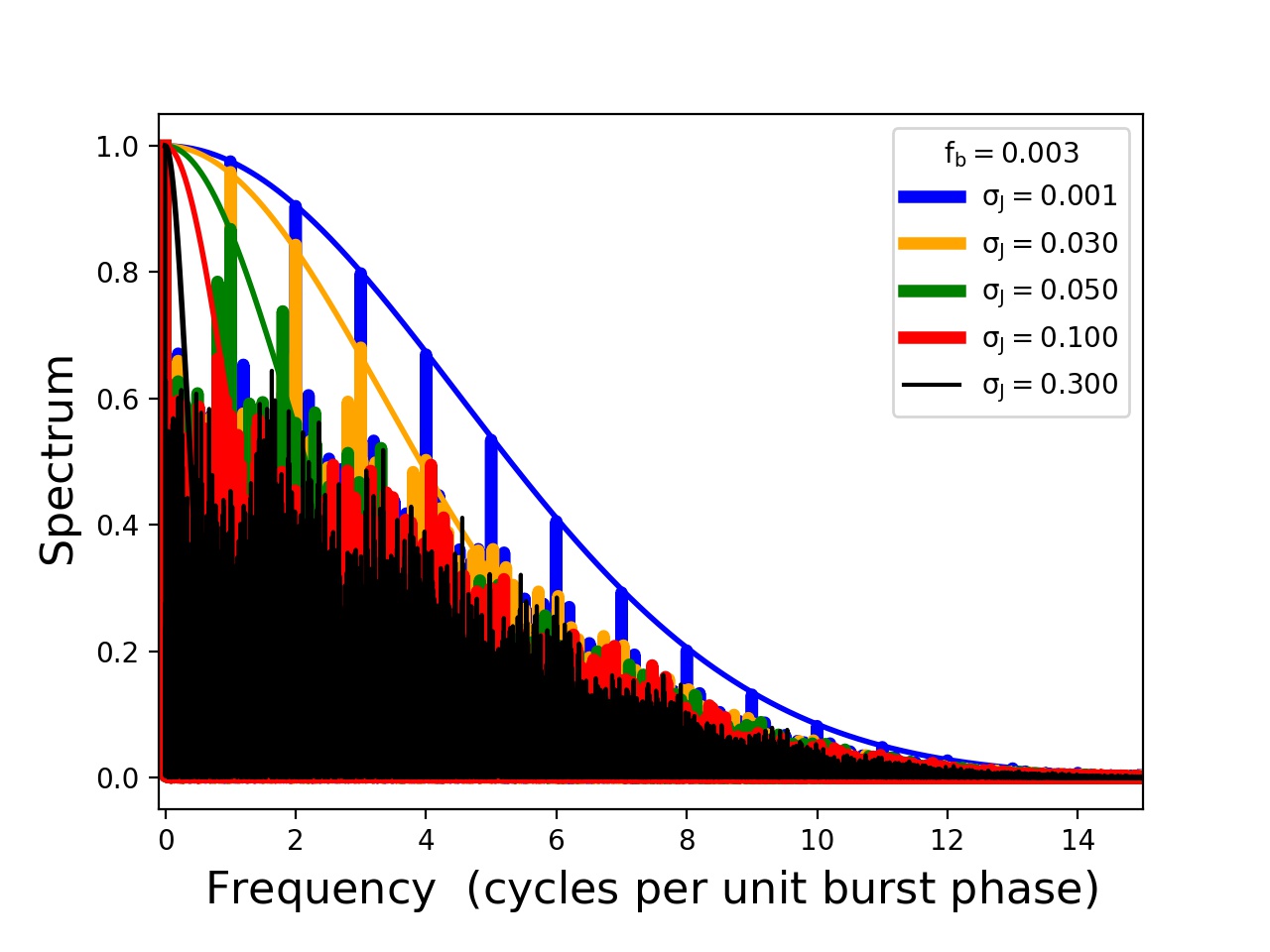} 
   \caption{Example power spectra for burst time series with different amounts of rms phase jitter,
   $\sigma_{\rm J}$, as indicated in the legend and shown with different colors.   Vertical bars are spectral lines and their amplitudes follow the envelope functions   $\propto \Jitt^2(f) \vert \FTA \vert^2$.  Spectra are shown
   for two burst fractions: $\fb = 0.1$  (left panel) and $\fb = 0.003$  (right panel).
   \label{fig:burst_power_spectra}
   }
\end{figure}

 Figure~\ref{fig:burst_power_spectra} shows spectra for $\Nb=5000$ and for  five values  of rms phase jitter $\sigma_{\rm J}$ and two values of burst fraction $\fb$.   For $\fb = 0.1$ (left panel), the attenuation of spectral lines with increasing rms jitter can be seen, with complete quenching for $\sigma_{\rm J} = 0.3$.    The right-hand panel shows the much larger continuum part of the spectrum that results from a smaller burst fraction, $\fb = 0.003$.      
 
While realistic cases invariably involve additive radiometer noise and radio frequency interference, detection of spectral lines is hindered even without these complications.
 They must have significant amplitudes relative to fluctuations  in the continuum part of the spectrum, which  are exponentially distributed if  $\Nb\gg 1$ bursts contribute.  

 Without any smoothing of the spectrum, the spectrum has an exponential probability density function (PDF) with the mean and rms at any frequency $f$ both equal to the mean spectrum of Eq.~\ref{eq:spec}.  For a detection threshold $\Smin$,    statistical variations lead to a false-positive probability 
$\pFP = \exp(-\Smin / \Smean) $ for each of  $N_f$  spectral values, implying
  a mean number of  false positives $\sim\pFP N_f$. 
A time series of length $\sim \Nb / \delta\phi$  samples, where $\delta\phi <  W_{\rm A}$ is the sample interval in phase units, yields $N_f > \Nb/2W_{\rm A}$.  
Spectral line detection then requires $R_{\rm L} (f) > \ln N_f$ or a mean number of actual bursts,
$\fb\Nb > (1 + \mb^2 - \fb) (\ln  N_f) / \Jitt^2(f)$.

  For a continuous time series spanning a few hours  with millisecond sampling, $N_f \sim 10^7$, specifying
 no more than one false positive implies a threshold $\Smin(f) / \Smean = \ln N_f \sim 16$.    
  Assuming a modulation index $\mb \sim 1$ and no jitter ($\Jitt = 1$), line detection requires  $\Nb\fb \gtrsim 32$ bursts with non-zero amplitudes to distinguish  spectral lines from the continuum.
 Phase jitter increases the requirement  by a factor $e^{+(2\pi \ell\rmsj)^2}$. 
The fundamental spectral line  ($\ell = 1$)   disappears rapidly as  $\sigma_\phi$  increases for fixed $\Nb$,
requiring   about 50 bursts
for $\rmsj = 0.1$~cycle and a much larger $10^3$ bursts for $\rmsj =  0.3$~cycle. 
Higher harmonics ($\ell \ge 2$) require many more bursts to be detectable.

The origin of jitter may be similar, but more extreme, than is seen from pulsars
where the emission beam integrated over many bursts is  wide but is instantaneously  
luminous in only a fraction of its overall solid angle.  Multiple beams are another possibility.  However, very wide beams lower the influence of beam wobble from precession because they are visible for a greater fraction of the precession cycle and  inconsistent with the 25 to 50\% duty cycle windows in which bursts are seen.    An alternative explanation is to associate phase jitter with retardation and rotational aberration that varies between bursts.   These effects can produce phase variations $\lesssim 1/\pi$ that are sufficient to hide spectral lines in the spectrum. 

\subsubsection{Distribution of Wait Times  Between Bursts}

The wait time or wait phase $\Delta\phi$  between contiguous {\it detected} bursts is another diagnostic for the periodicity or characteristic spacing of bursts \citep[][]{2021Natur.598...267L}.   For strictly periodic bursts that are all detectable, the PDF of phase separations $\Delta\phi$
is a delta function, $\pdfdphi(\Delta\phi) = \delta(\Delta\phi - 1)$.  When some bursts are too weak to be detected or have null amplitudes but are still strictly periodic, $\pdfdphi$ contains additional delta functions at integer phase separations $\ell$ whose amplitudes depend on the probability of having gaps of various lengths.   
For the simple case where the amplitude in each spin period is either a null or a detectable burst, the PDF is 
$\pdfdphi(\Delta\phi) = (1-\fnull) \sum_{\ell=1}^{\infty}\fnull^{\ell -1}\delta(\Delta\Phi-\ell)$, which follows from modeling transitions between bursts and nulls as a Markov process \cite[e.g.][]{2013ApJ...775...47C}.   For $\fnull \to 1$, the PDF is spread over a wide range of  $\ell$ extending to a multiple of   $\sim 1/(1-\fnull)$.   
These delta functions are blurred by  convolution with the PDF for phase jitter, $f_J(\Delta\phi)$, which becomes significant  for $\rmsj \gtrsim 0.3$~cycles, causing the periodicity to be masked  in the wait-phase distribution. 

Figure~\ref{fig:waithist}  shows distributions of wait phases vs. $\rmsj$ (left panel) and vs. $\fnull$ (right panel)
that illustrate how the periodicity is erased in the distribution for $\rmsj \gtrsim 0.3$~cycles and how the peak of the broad distribution shifts to larger wait phases as $\fnull$ increases.     

The wait-time distribution from the large FAST sample comprising $\sim 1600$  bursts \citep[][Figure 3]{2021Natur.598...267L} shows no evidence for periodicity even on individual days where the burst rate exceeds 100~h$^{-1}$.  It does show a secondary feature that peaks at a time separation $\sim 3$~ms that is due to substructure in some of the  bursts.  The primary feature shifts to larger time separations when the detection threshold is raised, as expected for randomly spaced bursts. 

The absence of an obvious periodicity in bursts from any of the repeating FRBs therefore does not imply the absence of an underlying physical periodicity in the source.   
However,  the maximum of the wait-time distribution can be used to put an upper bound on the spin period of the source.   If the periodicity is hidden primarily by phase jitter, the upper bound is only about ten times the period while nulling may yield a significantly larger upper bound, e.g. ten to 100 times the period.}{By inspection of Figure~\ref{fig:waithist}, conservative upper limits on any period can be estimated from the peak of the wait-time distribution.   If the periodicity is hidden primarily by phase jitter (left panel)  the upper bound is  about ten times the period  (based on the maximum of the distribution for $\rmsj \gtrsim 0.3$; note that the color scale is logarithmic) while nulling  (right panel) yields a  larger upper bound, e.g. ten to 100 times the period.

\begin{figure}[t] 
   \centering
   \includegraphics[width=0.475\figwidth]{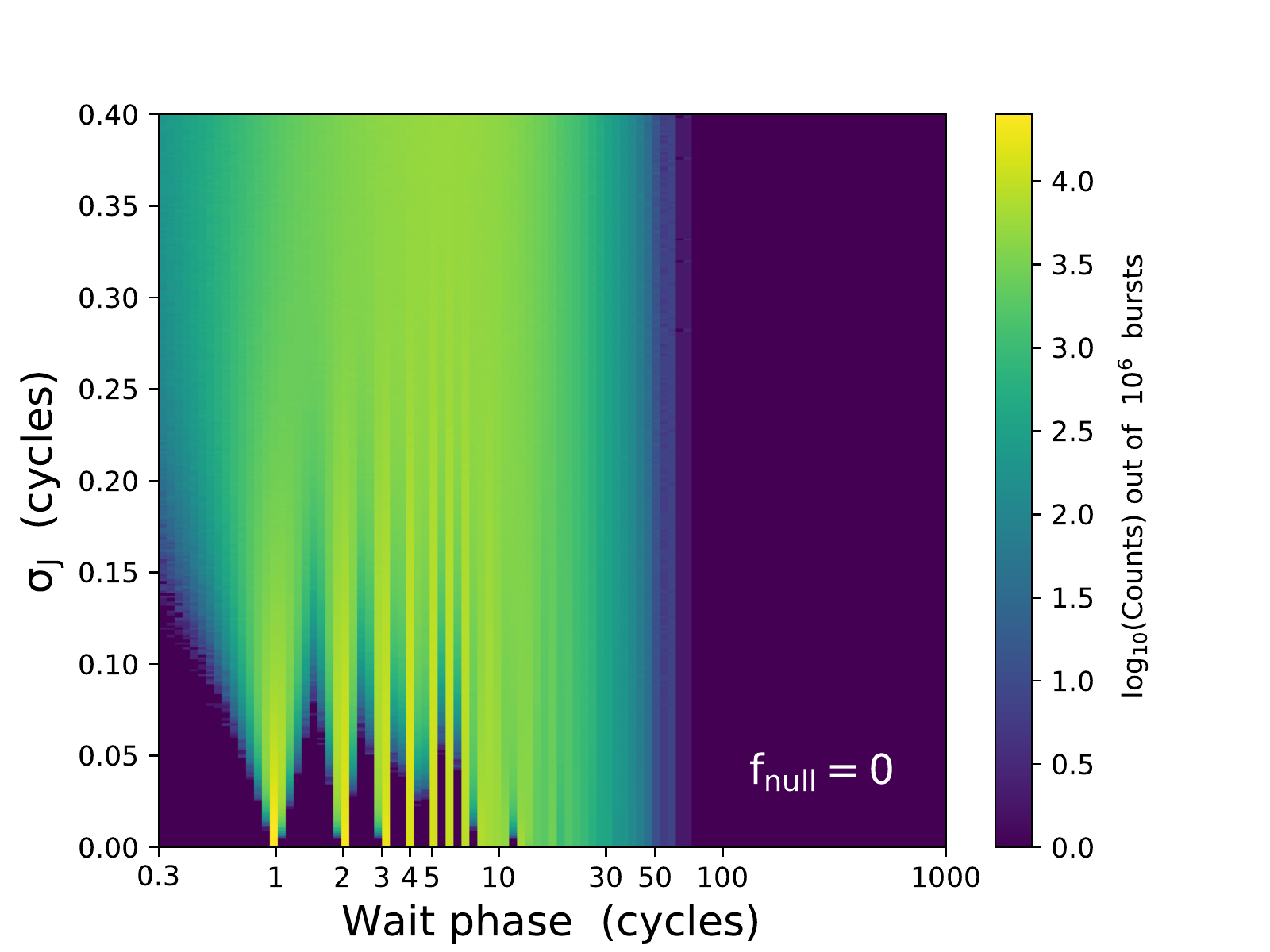} 
   \includegraphics[width=0.475\figwidth]{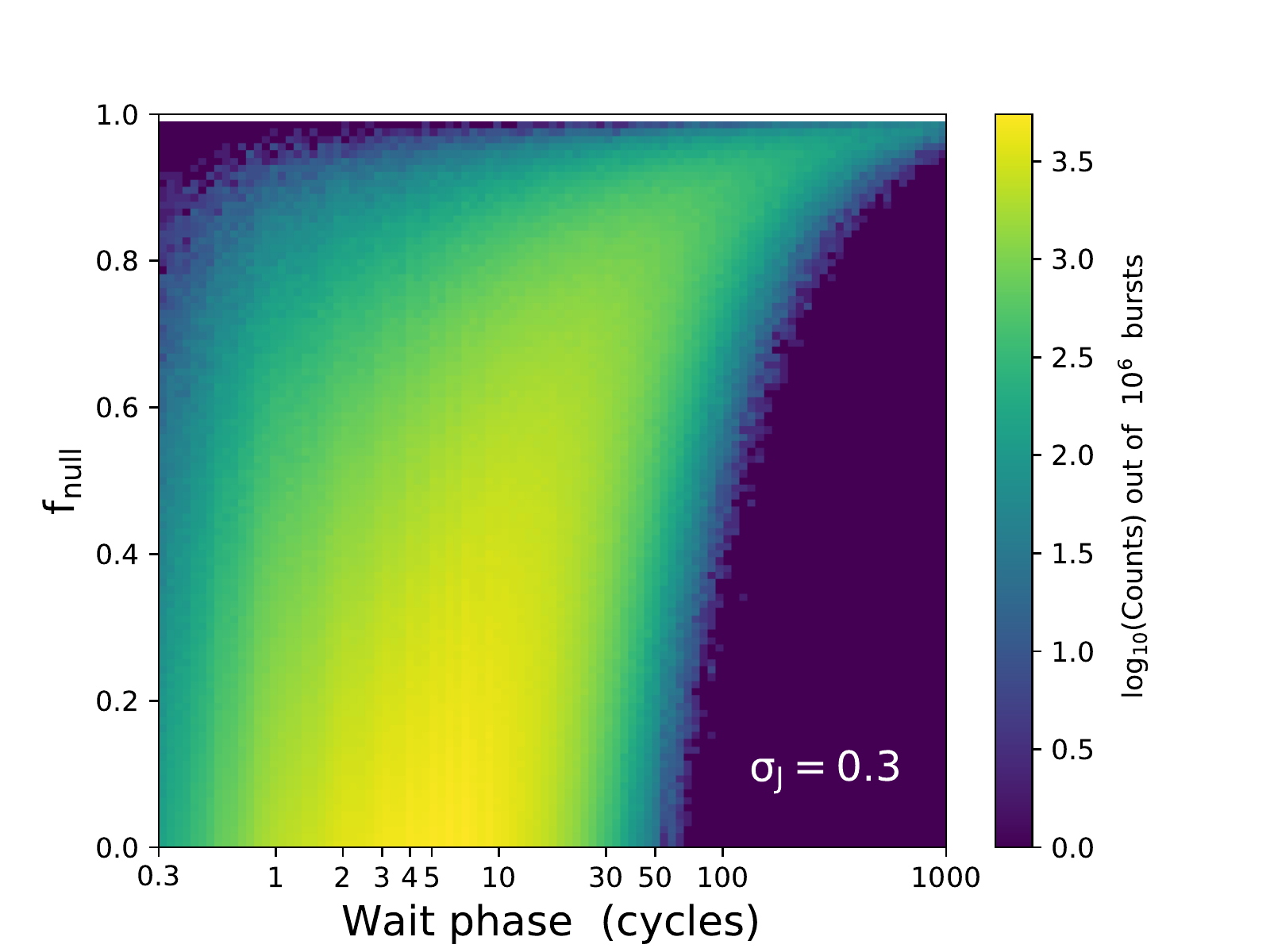}
   \caption{Histograms of the  phase difference between adjacent bursts that exceed a detection threshold. 
   They result from simulations of $10^6$ bursts with a skewed log-normal amplitude distribution and a detection threshold equal to the mean amplitude. 
  The color scale indicates the logarithm of the number of counts.
   Left: histograms vs. the RMS phase deviation from perfect periodicity $\sigma_{\rm J}$ where there  are no null pulses.
   Right: histograms vs. nulling fraction for a fixed RMS phase jitter, $\sigma_{\rm J} = 0.3~{\rm cycles}$.
  Preferred spacings at integer numbers of cycles are seen for small phase jitter but these vanish for
  $\sigma_{\rm J} \gtrsim 0.3$~cycles. For larger nulling fractions, the histograms shift progressively further to a larger mean value of the waiting phase. }
   \label{fig:waithist}
\end{figure}

\subsection{Periodicity Detection in Long Burst Sequences}

Burst sequences over periods of days to years require phase modeling to remove what are likely to be large contributions from the spindown torque and from the systematic orbital and precession terms
in Eq.~\ref{eq:phimodtotal} and \ref{eq:phimodfit}.  
If the modeled phase $\phim$ is accurate,
 evaluation at the measured arrival times gives zero residuals  in the absence of stochastic contributions from $\phisn, \phij$ and $\phin$. However,   the fractional part of $\phim(t_j)$ is generally  nonzero due to these  and any other unmodeled terms, giving  residuals 
 $\delta\phi(t; \thvec) =  \phim(t; \thvec) - {\rm int} \{\phim(t; \thvec)\}-1/2$ that are constrained to  the interval $[-1/2, 1/2]$.
Letting $\thvec$  represent all  model parameters for $\phim$, 
we define a  detection statistic as a  sum of phase factors dependent on $\thvec$
(with weights $w_j$ that sum to unity),
\be
D(\thvec) 
= 
\sum_{j=0}^{\Nb-1} w_j e^{2\pi i \delta\phi(t_j;\thvec)} .
\label{eq:detstat}
\ee
A perfect phase model  yields $D(\thvec)  = 1$ in the absence of the stochastic terms but when they are present,  
\be
D(\thvec)_{\rm max} 
= \etatotal 
= \langle e^{2\pi i\delta(\phisn + \phij + \phin)} \rangle .
\ee
If the  stochastic terms have large combined variance,  ${\rm Var}[\phi(t_j)] \gg 1 $, 
the resulting   uniformly distributed residuals in 
$[-1/2, 1/2]$, yield a mean $\langle D\rangle = 0$ and rms $\sigma_{\rm D} \simeq \ND^{-1/2}$. 
The magnitude $\vert D\vert$ is biased by a positive mean value $\sim \sigma_{\rm D} $
and has a Rayleigh PDF while $\vert D\vert^2$ is distributed as a one-sided exponential PDF. 

Periodicity detection requires all three of the stochastic phases to have  small variances $\ll 1$~cycle$^2$
individually as well as for their sum.  Then we can factor
$\etatotal$ into the product of individual form factors, giving a maximum
\be
D(\thvec)_{\rm max} =  \Jitt\, \etan \, \etasn(T).
\ee
To illustrate, consider a perfect model and phase jitter alone, which gives $\langle D \rangle = \Jitt$.  
We use the same jitter form factor as before in Eq.~\ref{appeq:Jitt}  (but with $f=1$).  
  Requiring $\langle D \rangle \gg \sigma_{\rm D}$ and using the burst fraction $\fb$ defined earlier, a data set needs to span $N = T / \Pspin$ spin periods and satisfy 
$N \fb \gg  \eta_{\rm J}^{-2}$, essentially the same constraint as from the power spectrum. 

The measurement noise form factor for Gaussian statistics is similarly
$
\etan = \langle e^{2\pi i \phin} \rangle = e^{-2(\pi \sigma_{\phin})^2}.
$
The form factor for correlated spin noise involves the phase structure function, 
$\SFphispin(t_i, t_j) = \langle [\phisn(t_i) - \phisn(t_j)]^2\rangle$ \citep[e.g.][]{1985ApJS...59..343C}.    Generally $\phisn$ has nonstationary statistics and $\SFphispin$ depends separately on $t_i$ and $t_j$.  Factoring out the arrival time of the first burst
and assuming  $\phisn$ also is a Gaussian random process, the  form factor is
\be
\etasn 
	=   \Nb^{-1}
	    \sum_{j=0}^{\Nb-1} \left\langle e^{2\pi i [\phisn(t_j) - \phisn(t_0)]}\right\rangle 
\longrightarrow 
T^{-1} \int_0^T d\tau\, e^{-2 \pi^2 \SFphispin(\tau)},
\label{eq:etasn}
\ee
where the last expression applies in the continuous limit over a data span of length $T = t_{\Nb}-t_0$.
As we show next,  the spin noise variance  and structure function grow as  power laws in $T$. 
For young magnetars and data spans of days or more,  the integrand in Eq.~\ref{eq:etasn} could be small for most of the interval $[0, T]$, yielding $\etasn \ll 1$. 

\begin{figure}[ht] 
   \centering
   \includegraphics[width=0.6\figwidth]{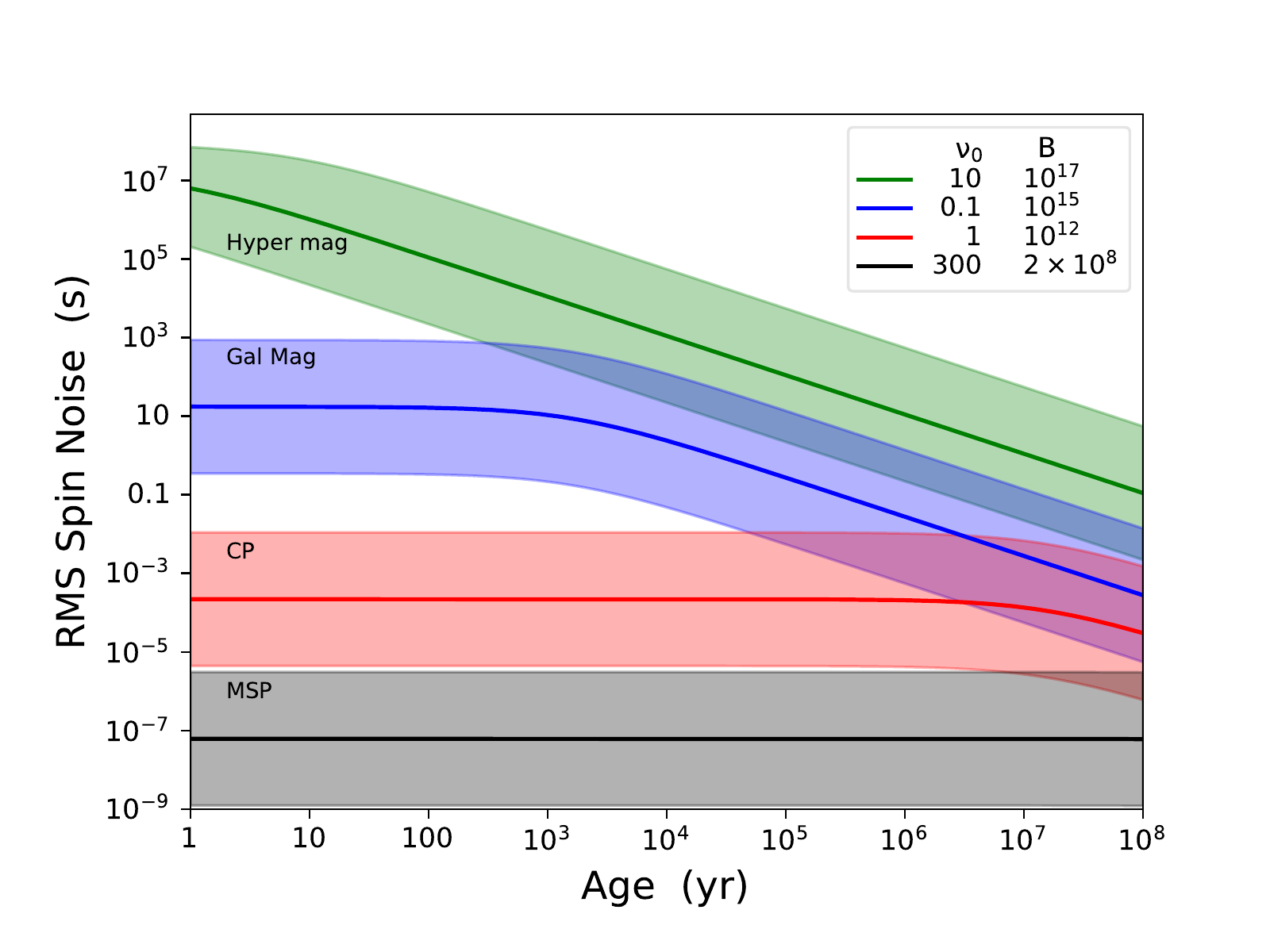} 
   \caption{
   RMS spin noise  vs. elapsed time since birth with an initial spin rate $\nu_0$ (indicated in the legend). The colored bands correspond to four classes of neutron stars and show the predicted RMS using the scaling law in Eq.~\ref{eq:sigt}, including uncertainties in the parameters.  The vertical range of the bands is dominated by the strong $T$ dependence because values 
   $T = 0.1$~yr and 10~yr on the lower and upper boundaries, respectively, have been used to show the  nonstationarity of the spin noise.   The four classes of neutron star include, in order of increasing surface magnetic field (also  in the legend),  millisecond pulsars (MSPs), canonical pulsars (CPs),  Galactic magnetars, and `hyper' magnetars with very large fields.   Initial spin rates $\nu_0$ are shown for the four classes.  For the first three classes, the specified magnetic field is
   the surface field, which we assume is dipolar in form.   For the hyper-magnetar, the surface dipole field is 10\% of the total field of $10^{17}$~G. 
   }
   \label{fig:rms1}
\end{figure}

\subsection{Spin Rate Variations}
\label{sec:spin}

Spin noise in pulsars typically appears as a `red' stochastic process having a  steep spectrum $\propto f^{-x}$ with $x \simeq 4$~to~6 and in some cases is due to resolved step functions in $\nu$ or $\nudot$
\citep[e.g.][and references therein]{1995MNRAS.277.1033D}. 
They may result from a combination of effects internal to the NS and in its magnetosphere.  Glitches are much larger events that may occur very frequently in young magnetars.    
Without elaborating on the underlying physics here, we simply extrapolate spin noise to young, high-field objects using  the scaling law for spin noise from pulsars  as a means for assessing its possible role in periodicity detection.  

In Appendix~\ref{app:spinnoise}, 
we express the  scaling law for spin noise in terms of the dipolar field   by making use of  the spindown rate,
$\vert\dot\nu_{-15}\vert \simeq 10^{6}\,{\rm Hz~s^{-1}}\, B_{\rm d_{15}}^2 \nu^3$
 for magnetic dipole radiation with a  fiducial dipole magnetic field strength $B_{\rm d} = B_{\rm d_{15}} 10^{15}$~G and a typical moment of inertia of $10^{45}$~g~cm$^{-2}$.  For a data span length  $T$  in years, the rms spin noise is
\be
\sigt 
\simeq
 1326^{+870}_{-519}  \, s  \times \nu^2 \, B_{\rm d_{15}}^{2.3}\, T^{1.7} .
\label{eq:sigt}
\ee

Figure~\ref{fig:rms1} shows $\sigt$~vs. age for different classes of NS, from millisecond pulsars (MSPs) to `hyper magnetars' with very strong fields.   The bottom and top of each colored band corresponds to data spans of length 0.1 and 10~yr respectively.
The two top cases for $10^{15}$ and $10^{17}$~G surface fields have very large RMS values that will mask the spin periodicity on year-like time scales.   In Paper I, our assessment of NS fields and spin dynamics suggests that surface field strengths are more likely to be closer to $10^{15}$~G  than to larger fields. 

For a year long data span (corresponding to the middle line plotted in each colored band in Figure~\ref{fig:rms1}),  the rms spin noise is clearly much larger than one cycle of phase for a one second period, so compensation for smooth spin down in the analysis of a long FRB time series is insufficient to identify a coherent periodicity if the number of bursts is sparse and they are widely spaced.
The data span length that corresponds to one cycle of phase variation ($\nu\sigma_t = 1$) from spin noise is
typically a few days for nominal parameters,
\be
T_1 \simeq 5.3^{+1.8}_{-1.4}\ {\rm d} \times \nu^{-1.76} B_{\rm d_{15}}^{-1.35}. 
\ee
Bursts   separated  by more than $T_1$ will not allow identification of a fast periodicity using any of the methods discussed above because   the form factor for spin noise would  be very small $\etasn \ll 1$.  Conversely, short data spans with $T \ll T_1$ with multiple bursts should allow detection of any periodicity. 

Any magnetic field decay  reduces the rate of spindown, allowing  large spin rates to be sustained for longer times.    The higher spin rate makes the spin noise larger than otherwise at a given age but the lower  field strength reduces the spin noise (cf. Eq.~\ref{eq:sigt}), so the net result is unclear.   Future  analyses on repeating FRBs may ultimately detect periodicities in short burst sequences, which will allow constraints on spin noise if burst sequences from the same objects become less coherent in sequences of days or longer.  This will provide information about the underlying physics of spin noise in young magnetars.  For now, however,  the absence of burst periodicities in short burst sequences must be explained by effect(s) other than spin noise, such as  phase jitter discussed above. 

\subsection{Unmodeled Systematic Variations}
\label{sec:unmodeled}

Deterministic contributions to the phase model from  precession (or orbital motion) will also inhibit detection of a fast periodicity if sparse burst time series cover multiple precession or orbital periods.      
Precession presents a greater challenge, in general, than orbital motion.  It perturbs arrival times in two ways: through wobble of the emission beam(s) relative to a non-precessing object and by inducing a cyclical torque resulting from the dependence of the torque on the spin-magnetic moment angle 
\citep[][]{1988MNRAS.235..545J,1991PhDT.........1B,1993ASPC...36...43C}.
  The beam wobble contribution ultimately depends on the shape of the emission beam and how it moves across the observer's direction.   In the special case of low-amplitude precession  with a simple, Gaussian-like emission beam, the precession perturbation may be nearly sinusoidal and mimic orbital motion \cite[e.g.][]{1990ApJ...348..226N}.  However,  triaxial precession combined with a complex beam shape will depart from a simple fitting function to arrival times.    
  
For young magnetars, we expect the phase contribution from the cyclic torque
$\Dphicyctorque$ to dominate beam wobble $\Dphiwobble$, as discussed in detail for triaxial precession in \S\ref{sec:timing}.  For small-amplitude precession
 $\Dphiwobble \sim \theta_{\rm p} / 2\pi  \ll 1$~cycle where $\theta_{\rm p}$
is the amplitude of  the change in the spin-magnetic-moment angle.  A simple scaling law gives 
$\Dphicyctorque \sim \nudot \theta_{\rm p} \Pp^2 / 4\pi^2 \sim \nu \theta_{\rm p} \Pp^2 / 8\pi^2\taus $
(Appendix~\ref{app:cyc_torque}).
 For a precession  period $P_{\rm p} = 10^6 P_{\rm p, 6}$~s and a characteristic spindown time  $\tau_{100}  = \nu / (2\dot\nu \times100\,{\rm yr})$, this gives $\Dphicyctorque \sim 2~{\rm cycles} \times  \theta_{\rm p} P_{\rm p, 6}^2 / \Pspin \tau_{100}$.    If uncorrected,  $\Dphicyctorque$  will mask the fast periodicity in  sparse burst sequences that span one or more precession periods.     In principle, precession can be fitted and removed to enable identification of fast periodicities, but the precession parameter space is potentially very large, as discussed in the next section.  Imperfect removal can yield timing residuals $\delta\phi(t)$ that are easily large enough to mask the spin periodicity through a form factor $\vert \langle e^{i\delta\phi(t)}\rangle\vert \ll 1$.   
Triaxial precession presents very different time dependences for the wobble and torque effects than low-amplitude precession from an axisymmetric star. 

We again comment on the role of decay of the dipolar magnetic field  in objects older than, say, 1~kyr.   The amplitude of the  cyclical torque $\Dphicyctorque \propto \nudot \propto B_{\rm d}^2$ decreases with field decay.  This effect, along with others discussed in Paper~I, allow the possibility that  burst periodicities hidden in young objects may ultimately manifest in older objects even for burst sequences that span multiple days or weeks. 

\section{Triaxial Precession}
\label{sec:triaxial}

Paper I gives a detailed description of the precession of a triaxial star and its consequences.  Here we define those quantities needed in our analysis of the observational manifestations of precession.

The triaxiality of the star is a consequence of magnetic distortion quantified by $\epsmag$, which determines  the principal components of the moment of inertia tensor, 
$I_i$,  where $i=1,2,3$.  The measure of triaxility used is
$e^2 \equiv I_3(I_2-I_1) / I_1(I_3-I_2)$, which is zero for an oblate, axisymmetric star.  
For unit vectors $\ehat_i$ along the principal axes, the angular momentum is directed along the unit vector $\lvh = \lvhh_i \ehat_i$ (summation convention implied) with components
\be
\lvhh_1=\Lambda\cn(\Phi), \quad 
\lvhh_2=\Lambda\sqrt{1+e^2}\,\sn(\Phi), \quad 
\lvhh_3=\sqrt{1-\Lambda^2}\,\dn(\Phi)
\label{eq:lvhsol}
\ee
where $\cn(\Phi), \sn(\Phi)$ and $\dn(\Phi)$ are Jacobian elliptic functions
\citep[e.g][]{1972hmfw.book.....A}.  A full precession cycle is  $\Phicyc = 4 F(\pi/2|q)$,  where $F(\pi|q)$ is the complete elliptic function of the second kind 
and $q = e\Lambda / \sqrt{1-\Lambda^2}$.
For $e^2 >0$, the phase for one precession cycle is $\Phicyc > 2\pi$. 
For an oblate star ($e^2=0$), the expressions simplify with $\cn \to \cos$, $\sn \to \sin$, and $\dn \to 1$
and  one precession cycle becomes  $\Phicyc=2\pi$. 

Figure~\ref{fig:precession_geometry} shows the precession geometry in a top view (left) and side view (right) in a frame fixed with the neutron star.    The angular momentum vector traces an ellipse in the $\lvhh_1$ - $\lvhh_2$ plane that becomes a circle for $e^2 = 0$.    The $\lvhh_3$ component also includes nodding motion during a precession cycle.    Another presentation of the geometry is given in Figure~1 of Paper I. 

 A key feature of free precession  is that the angle between the magnetic moment $\muhat$ and the instantaneous
spin vector along $\lhatvec$ varies over a precession cycle, causing a cyclical change in  the magnetic torque on the star that superposes with the mean torque. 

The precession phase $\Phi$ is related to spin phase and to the precession and spin periods by  
\be
\frac{d\Phi}{d\phi}
= \frac{2\epsmag\sqrt{(1-\Lambda^2)(1+e^2)}}{2+e^2}
= \left[ \frac{F(\pi/2 \vert q)} {\pi/2} \right] \frac{\Pspin}{\Pp}~.
\label{eq:dPhidphi}
\ee
For an axisymmetric star, $\Lambda = e^2 = q = 0$ and $\Pspin / \Pp = \epsmag$, the familiar ratio  of periods in terms of the star's ellipticity, $\epsmag$.

\begin{figure}[b!] 
   \centering
   \includegraphics[width=0.45\figwidth]{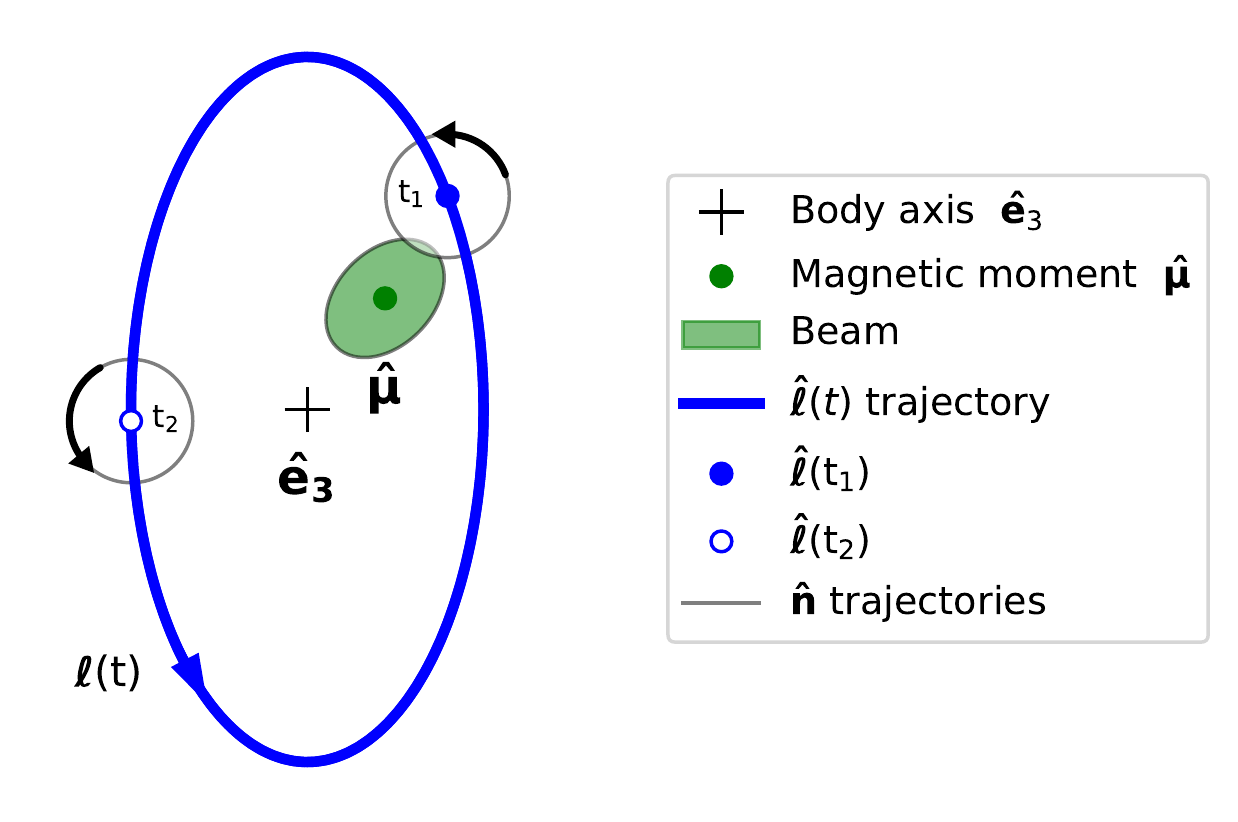} 
     \includegraphics[width=0.35\figwidth]{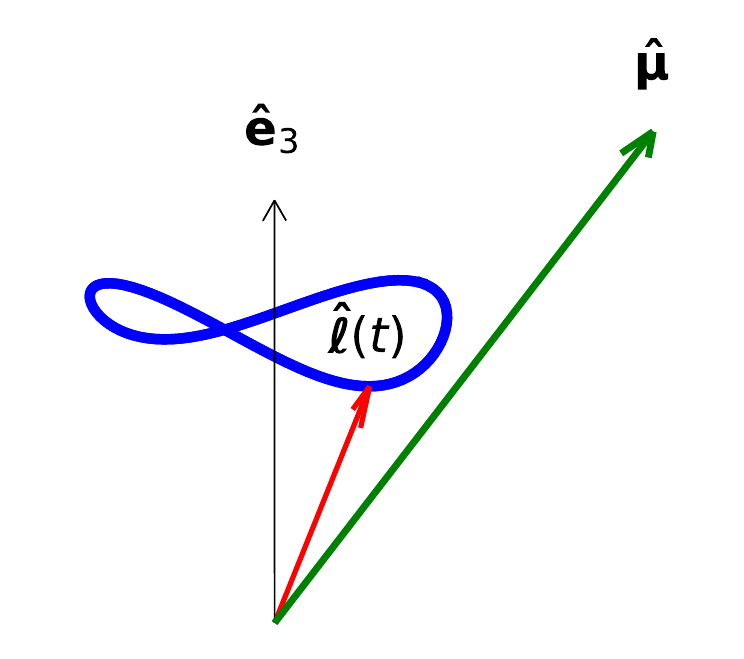}
   \caption{Precession geometry in the body frame.
   Left:  View looking down on the principal moment of inertia, 
   ${\bf \hat e}_3$.   In this frame the magnetic moment $\muhat$ is fixed and the example emission beam is shown as a    green ellipse centered on $\muhat$.  Generally it could  be oriented in some other direction. 
  The angular momentum makes an elliptical trajectory around ${\bf \hat e}_3$ and the observer's line of sight rotates around the instantaneous orientation of $\lhatvec$, as shown at two different times $t_{1,2}$ during the precession cycle. 
  Right: Side view for a slightly different case where $\muhat$ is outside the locus of $\lhatvec$.
   }
   \label{fig:precession_geometry}
\end{figure}

\section{The Beam Precession  Modulation Function}
\label{sec:bpmf}

To investigate the effects of precession, 
we define the observer's direction in the inertial frame relative to the source as 
\be
\nhat=\cosi\lvh+\sini\ehat_x~,
\ee
in a right-handed coordinate system $\ehat_x$, $\ehat_y$ and $\ehat_z=\lvh$. 
The emission beam is  fixed in the body frame  with a maximum in the direction given by the angles $\theta_{\rm b}$ and $\phi_{\rm b}$, 
\be
\bhat=b_1\ehat_1+b_2\ehat_2+b_3\ehat_3=\sintheta_b(\cosphi_b\ehat_1+\sinphi_b\ehat_2)+\costheta_b\ehat_3~.
\ee
For radio pulsars, the beam is tied to  the dipole axis $\muhat$, either nearly parallel to it for `polar cap' emission or skewed for high altitude `outer gap' emission.  Rotational aberration of course plays a role in the actual beam direction.

Figure~\ref{fig:precession_geometry} shows that the beam indicated by a green-shaded ellipse centered on $\muhat$  is intersected by the line of sight only at a subset of precession phases. 

Bursts may involve sweep of the beam across the line of sight or they may be temporal phenomena with a greater probability of detection when the beam points near $\nhat$. 
For specificity, we adopt a beam that is circularly symmetric around $\bhat$, so the  observed intensity (or burst probability)  is a function of $\bhat\dotprod\nhat$.  In particular, we use a Gaussian beam
with 1/e half width $\theta_{\rm e} = \theta_{\rm FWHM} / 2\sqrt{\ln 2}$, 
\be
\Bscr(\bhat\cdot\nhat) = e^{-2(1-\bhat\cdot\nhat)/ \theta_{\rm e}^2} .
\ee
Using expressions in \S~2.3 of Paper I (Eq.~50 - 54),  the dot product is 
\ba
& &\bhat\dotprod\nhat=\cosi[\sintheta_b(\lvhh_1\cosphi_b+\lvhh_2\sinphi_b)+\lvhh_3\costheta_b]
\nonumber\\& &~~~~~
+\sini\left[\cos\gamma\left(\sqrt{\lvhh_1^2+\lvhh_2^2}\costheta_b
-\frac{\lvhh_3\sintheta_b(\lvhh_1\cosphi_b+\lvhh_2\sinphi_b)}{\sqrt{\lvhh_1^2+\lvhh_2^2}}\right)+\frac{\sin\gamma\sintheta_b(\lvhh_2\cosphi_b-\lvhh_1\sinphi_b)}
{\sqrt{\lvhh_1^2+\lvhh_2^2}}\right]
\label{eq:bdotn}
\ea
after  rotating by  an Euler angle  $\gamma$ given by  integrating (Paper I),
\be
\frac{d\gamma}{d\phi}
=-1 - \KeL \frac{d\Phi}{d\phi}\frac{1}{[1+e^2\sn^2(\Phi)]} 
\label{eq:dgammadphi}
\ee
with $\KeL \equiv \sqrt{(1+e^2) / (1-\Lambda^2)}$
for the oblate case with $\Lambda\sqrt{1+e^2} < 1$ presented in Paper I, which we use for purpose of illustration. 
Integrating Eq.~\ref{eq:dgammadphi} and setting  the constant of integration to zero, the Euler angle,
\be
\gamma =  -\phi \left[ 1 + \KeL \dPdp G(\Phi, e^2)\right] ~,
\label{eq:gamphi}
\ee
 is  constrained to be in $[0,2\pi]$.
The second term in  square brackets  is much smaller than unity because 
$\dPdpinline \sim \epsmag \ll 1$ while $\KeL \sim \order{1}$ and the
dimensionless factor $G \le 1$, where
\be
G(\Phi, e^2) = \frac{1}{\Phi} \int_0^{\Phi} \frac{d\Phi' }{1+e^2 \sn^2(\Phi')}.
\ee
 
As the star rotates and precesses, $\bhat\dotprod\nhat$ maximizes at a spin phase that  varies slowly as a function of $\Phi$.  Irrespective of the beam model, the value of $1-\bhat\dotprod\nhat$ is a measure of the detectability of the beam for any particular pulse period: smaller 
values are more favorable for detection. 

We define  the {\it beam precession modulation function}  (BPMF) as the amplitude of the beam function when it makes its closest approach   to $\nhat$  during a spin period at a given precession phase.   It therefore
represents the window in precession phase 
$\Phi$ in which burst amplitudes are  maximized, on average.   It is determined by the precession geometry in concert with the beam shape of the emission,  
\be
\BPMF(\Phi) 
= \Bscr((\bhat\cdot\nhat)_{\rm max}).
\ee

Modulation functions are shown  for  three precession cases in Figure~\ref{fig:bpmf1}.  On the left are  line plots of the BPMF for different beam widths at  a fixed observer's inclination while the right-hand panels show variations for different  inclinations as well as with beam width.    
The top row is for an axisymmetric, oblate star ($e^2 = 0$).    The BPMF is double peaked for most inclinations with separations between peaks that depend on inclination. 
The middle and bottom rows are for triaxial precession with $e^2 = 10$.    The beam direction is in the $\hat e_1 - \hat e_3$ plane
 (i.e. $\phi_{\rm b} = 0$) for the middle row and skewed in the bottom row with $\phi_{\rm b} = 30$~deg.  Salient features include:
\begin{enumerate}
\itemsep -1pt
\item Larger beam widths, not surprisingly, yield  BPMFs that are large for a greater fraction of the precession cycle.  
\item  The BPMFs are bimodal or trimodal in most cases.    The spacing of the modes is generally non-uniform and depends on the inclination.
\item For $\phi_{\rm b} = 0$ the BPMF is symmetric about $\Phi/\Phicyc = 1/2$ but the symmetry is broken for $\phi_{\rm b} \ne 0$. 
\end{enumerate}

The multiple modes of the BPMFs need to be considered in any precession interpretation of the slow periodicities seen from \Rone\ and \Rthree.    
Current observational constraints  indicate that bursts occur quasi-periodically in windows with
duty cycles $\sim 25$~to~55\%.  
From the BPMFs shown in the figures,  it is clear that only a subset of geometries and triaxialities will 
match the observations, but with considerable leeway on parameters owing to the paucity of bursts. 
In particular,  the available data are not  informative of the burst rate {\it within} the precession window and do not disallow closely-spaced double peaks in the BPMF.   In addition, for some of the cases shown in the figures,  the observed spacings of   $\sim 16$ and 160~d could be  submultiples of the true precession periods.    What is more certain, however, is that the beam width cannot be larger than about 
20~deg in order to match the observed duty cycles of the slow periodicities.

The effects of precession depend on the magnetic field strength in several ways.  The deformation of the star and the misalignment of the spin and principal axis require large fields; these in turn affect the amplitude and period of precession.   If magnetic fields decay during the FRB-emitting phase of a magnetar, all of these aspects of precession will likely be reduced.    For example, precession periods will become longer and amplitudes smaller, producing smaller precession-driven modulations of burst sequences.  In other words,   slow periodicities should become slower and the precession phase windows in which bursts are seen will become wider.     That suggests that in R1 and R3,  the field strengths must be large if precession is responsible for the slow periodicities.

\begin{figure}[t] 
   \centering
   \includegraphics[width=\figwidthtwo]{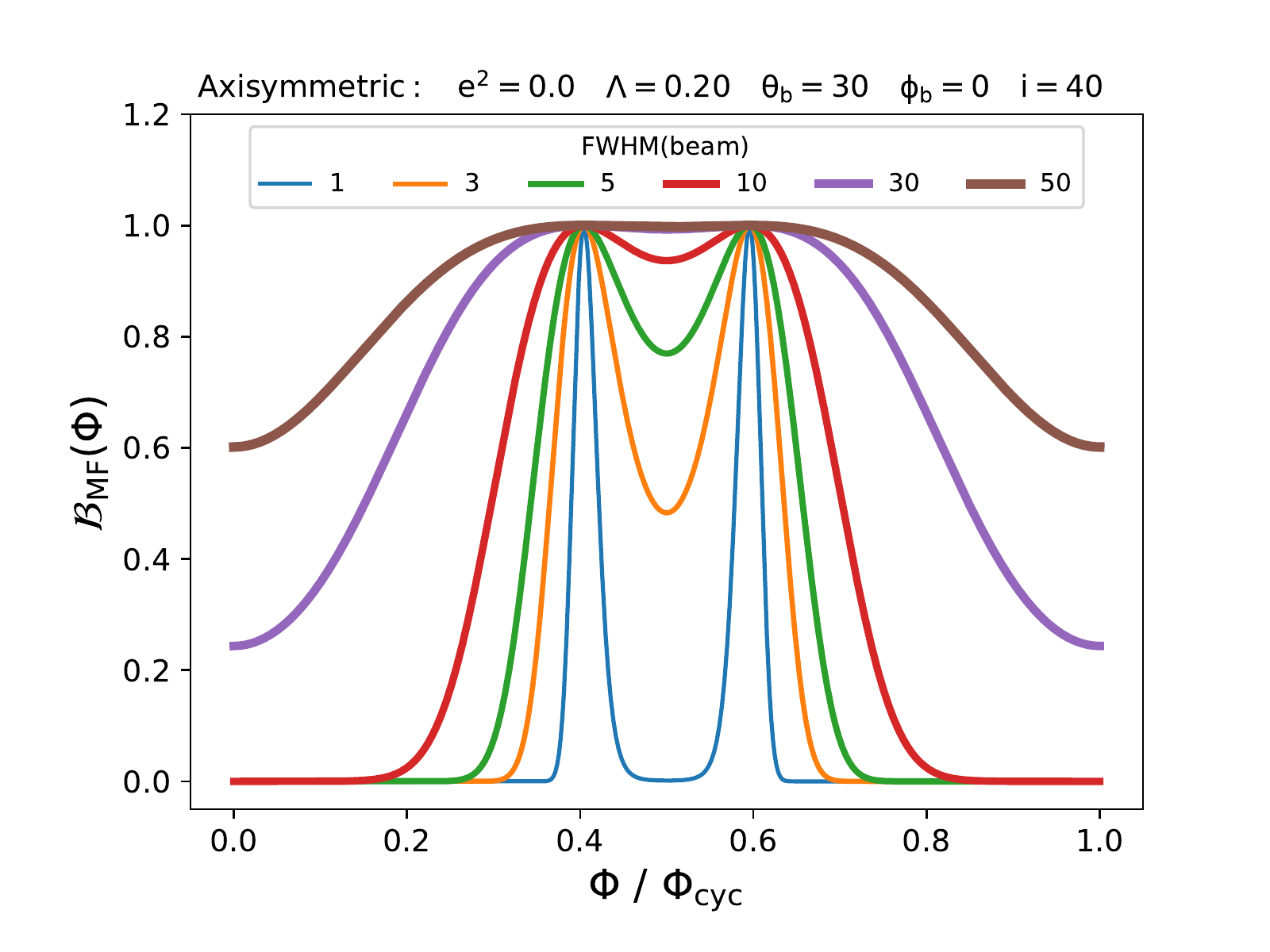}
\includegraphics[width=\figwidthtwo]{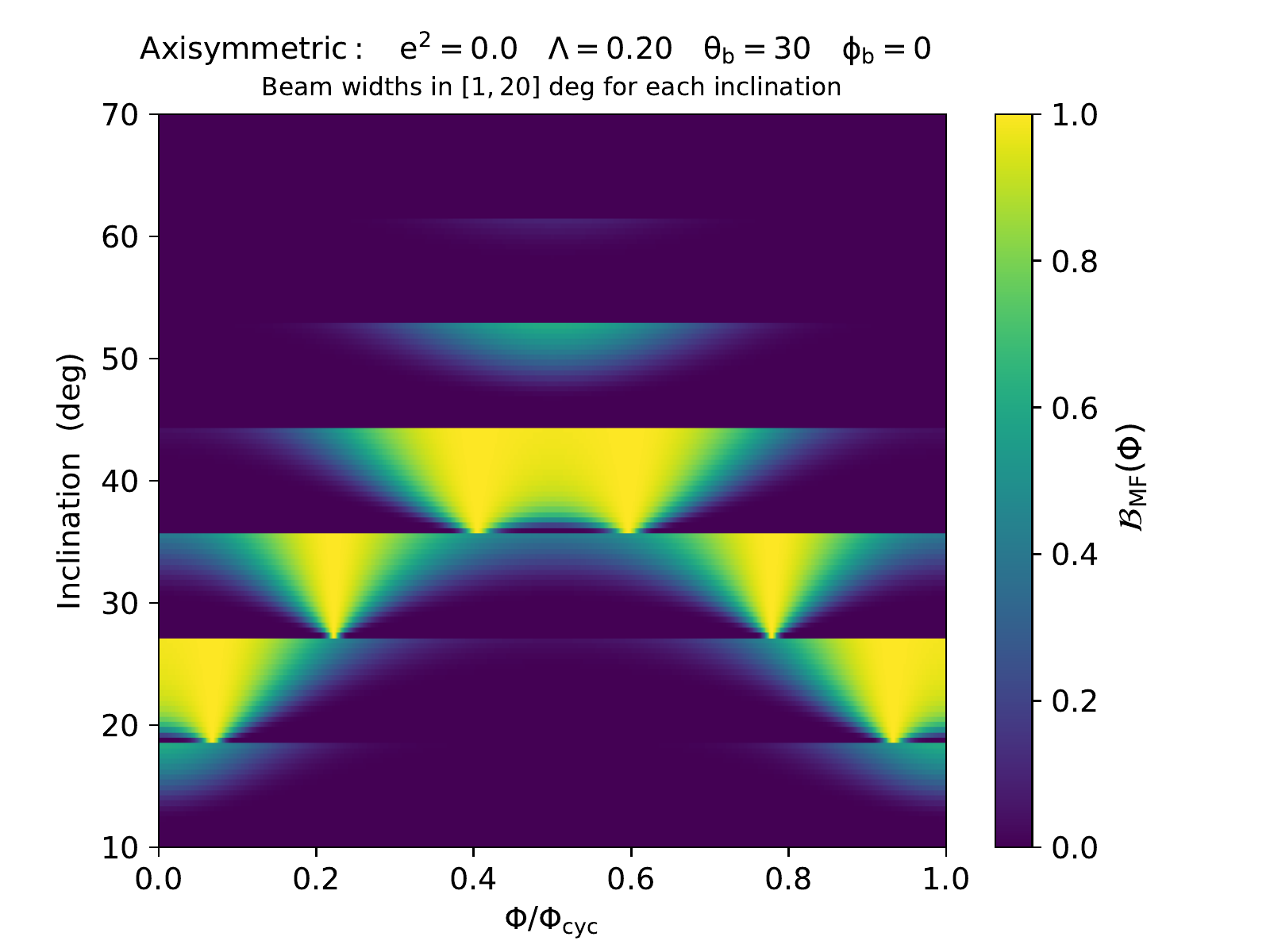}
\includegraphics[width=\figwidthtwo]{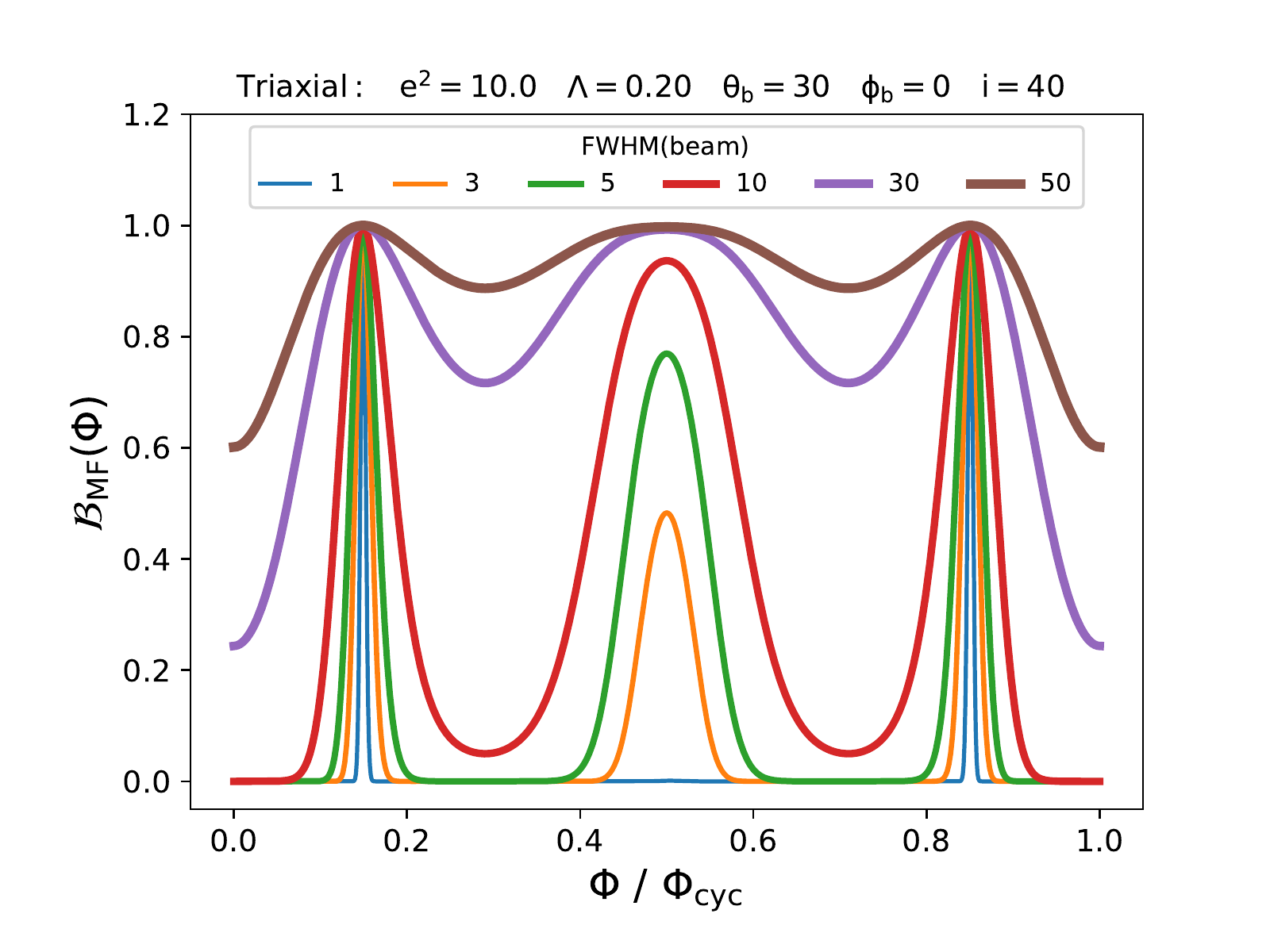}
\includegraphics[width=\figwidthtwo]{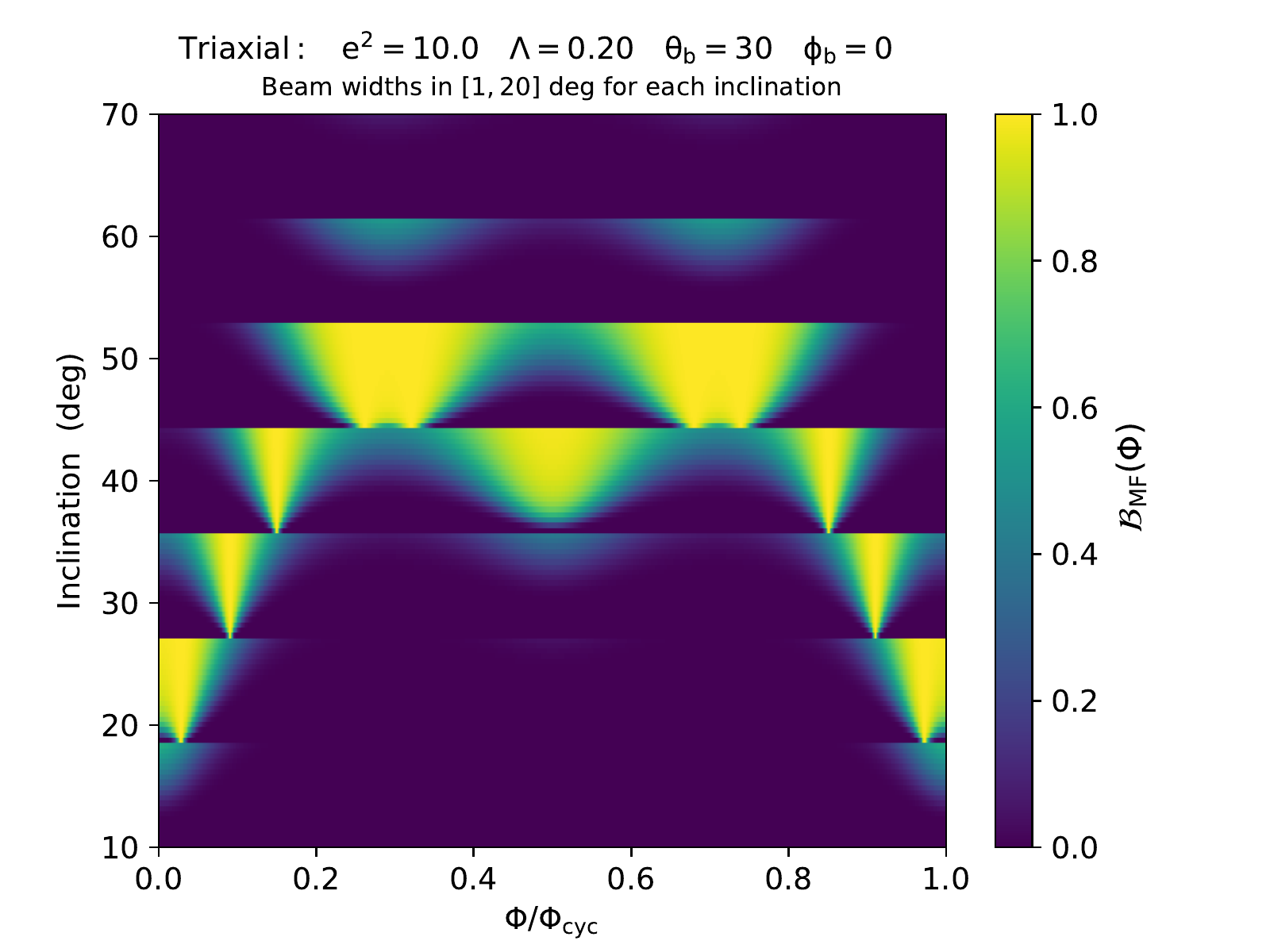}
\includegraphics[width=\figwidthtwo]{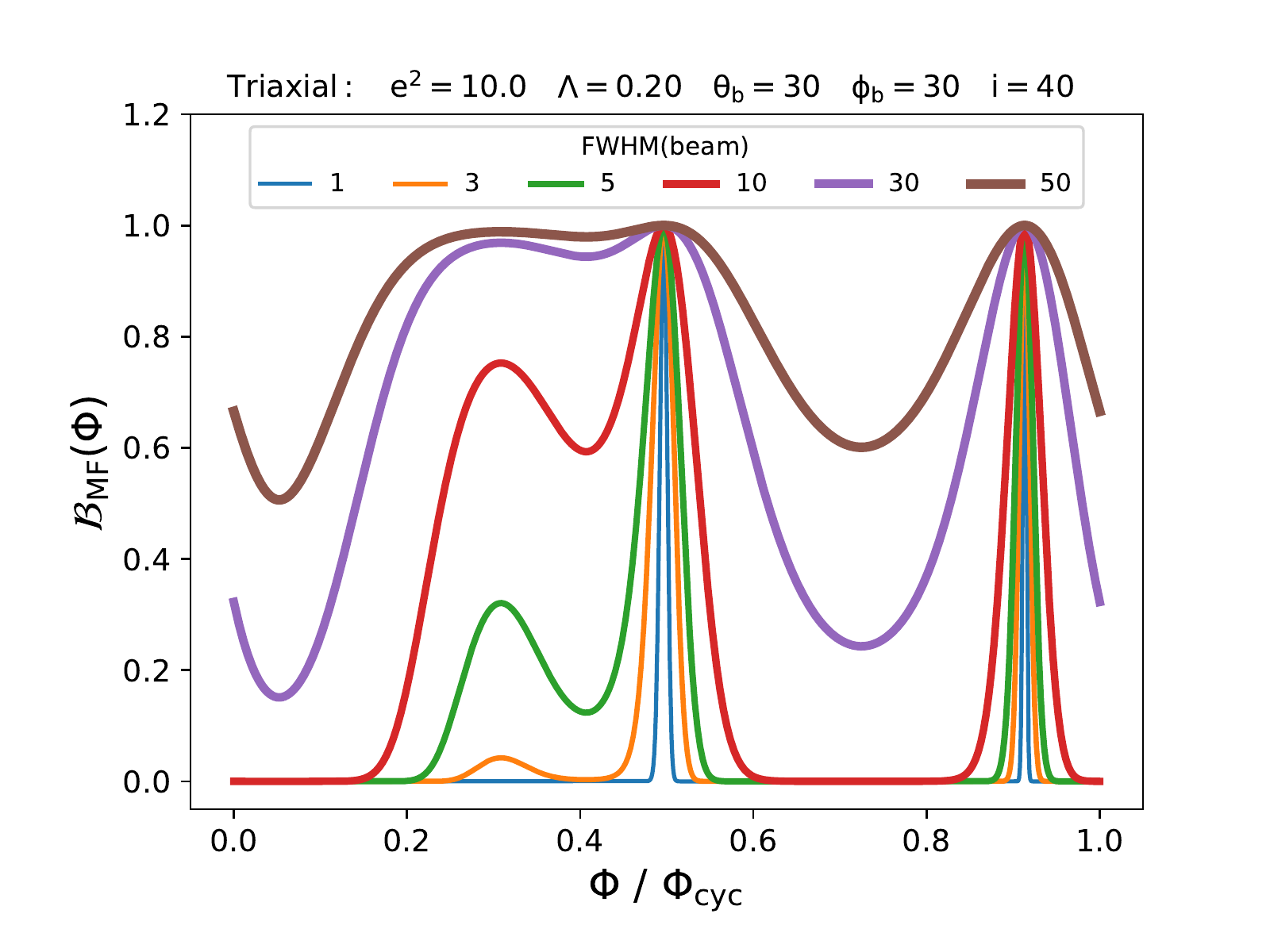}
\includegraphics[width=\figwidthtwo]{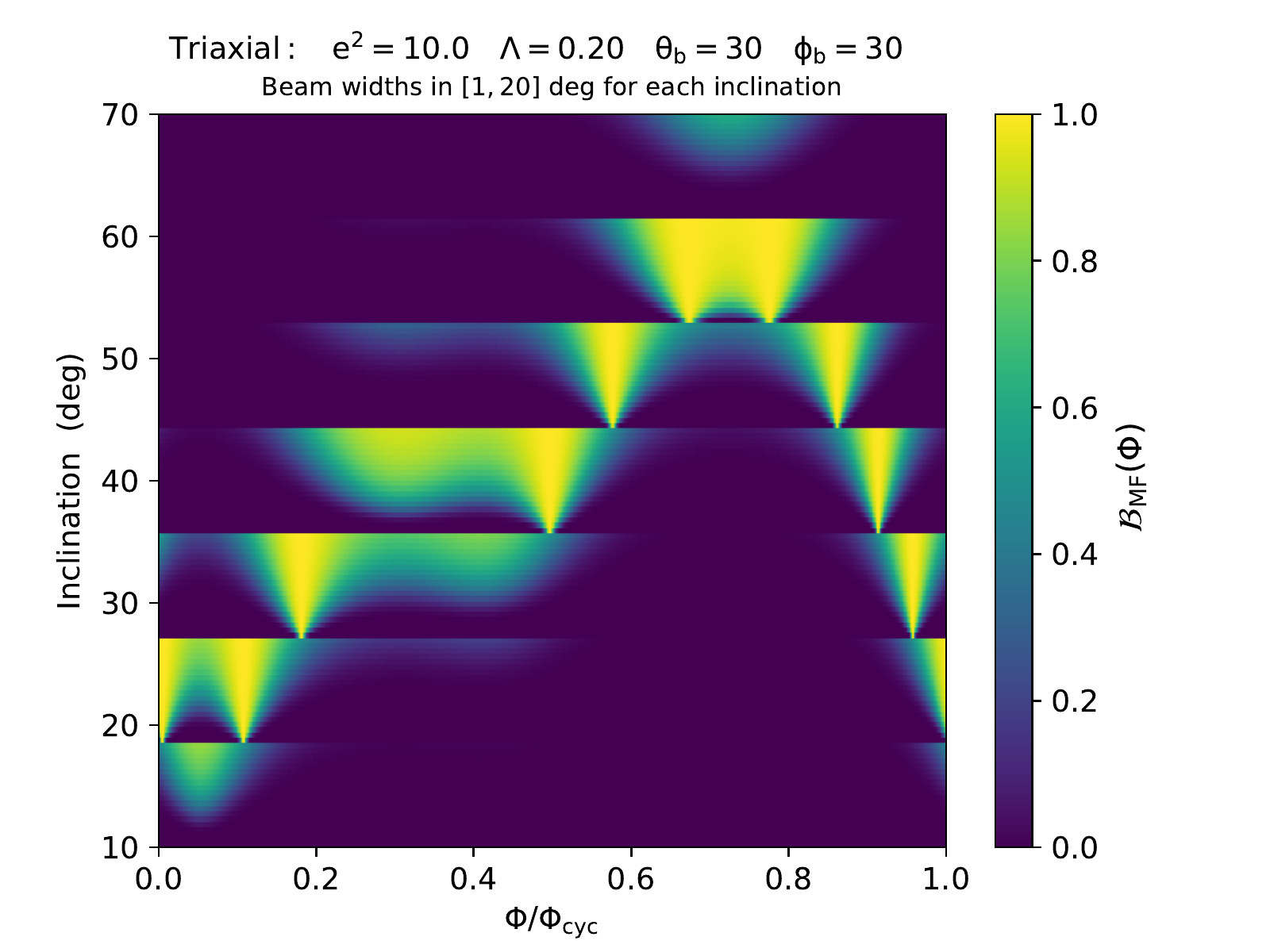}
\caption{The beam precession modulation function (BPMF) for different geometries.   
The left panel in each row shows line curves for different widths of Gaussian beams for a beam orientation $\theta_{\rm b} = 30$~deg. 
The right panel shows the BPMF amplitudes (color bar) for seven different  inclinations and for beamwidths from 1 to 20~deg for each inclination angle. 
Top: axisymmetric, oblate  precession with  $e^2=0$  and $\phi_{\rm b} = 0$. 
Middle: triaxial precession with  $e^2=10$  and $\phi_{\rm b} = 0$.
Bottom: triaxial precession with  $e^2=10$  and $\phi_{\rm b} = 30$ deg.
   }
   \label{fig:bpmf1}
\end{figure}

\section{Burst Timing Variations from Precession} 
\label{sec:timing}

In this section we consider arrival time perturbations that arise from precession.  We ignore spin noise here and though we include a spindown torque, we do so only by considering how the torque varies over a precession cycle. The steady spindown from the torque is ignored  but it is highly likely that it requires consideration in searching for a fast periodicity in burst sequences spanning multiple  precession cycles.     In addition to the cyclical variation in torque, we also analyze wobble of the emission beam. 

\clearpage
\subsection{Beam Wobble and Peak Intensities}
\label{sec:wobble}

Here we calculate the pulse phase departure from what it would be under strictly uniform periodicity without precession or spindown.  We consider the effects of precession of the radio beam over a single precession cycle, $0 \le \Phi \le \Phicyc$.
We give a short summary here while details can be found  in Paper I.

In order to track the detectability of the beam, we maximize the dot product by solving
\be
\frac{d\bhat\dotprod\nhat}{d\phi}= A(\Phi) + B(\Phi) \cos\gamma + C(\Phi) \sin\gamma = 0,
\label{bnzero}
\ee
where $A, B$, and  $C$ are functions of $\Phi$ and the precession parameters using terms collected from Eq.~\ref{eq:bdotn}.
This yields 
\be
\cos[\gamma(\Phi) - \Psi(\Phi)]= \Sigma(\Phi) ,
\label{eq:gPS}
\ee
where $\Sigma(\Phi) = A / \sqrt{B^2 + C^2}$ and $\Psi = \arctan C/B$; note that the signs of $B$ and $C$ need to be considered to find the proper value of $\Psi$. 

To calculate the phase residual from wobble of the beam caused by precession we solve Eq.~\ref{eq:gPS} to obtain  the Euler angle $\gamma_1(\Phi)$ over one precession cycle, $0 \le \Phi \le \Phicyc$.   
Choosing the solution that maximizes $\bhat\dotprod\nhat$,
the sequence of values vs. $\Phi$  has a wraparound of $2\pi$ at some $\Phi$ that we remove, 
 yielding a sequence $\gamma_{1u}$ that does not have these discontinuities, where the subscript `u' denotes
that $\gamma_1$ has been ``unwrapped'' 
 and the spin phase due to beam wobble is, 
\be
\Dphiwobble(\Phi) = 
    - \left\{ \gamma_{\rm 1u}(\Phi)  +  \KeL \Phi \left[ G(\Phi, e^2) - G(\Phicyc, e^2) \right] + 2\pi\Phi/\Phicyc \right\} .
\ee

Figure~\ref{fig:toavariations} (left panel) shows an example  dot product $\bhat\cdot\nhat$ vs. precession phase and the corresponding  spin phase perturbation from  beam wobble, 
$\Dphiwobble$.   The case shown is for a triaxial star and a beam at azimuthal angle $\phi_{\rm b} = 30$~deg. 

\begin{figure}[ht]
\centering
 \includegraphics[width=\figwidthtwo]{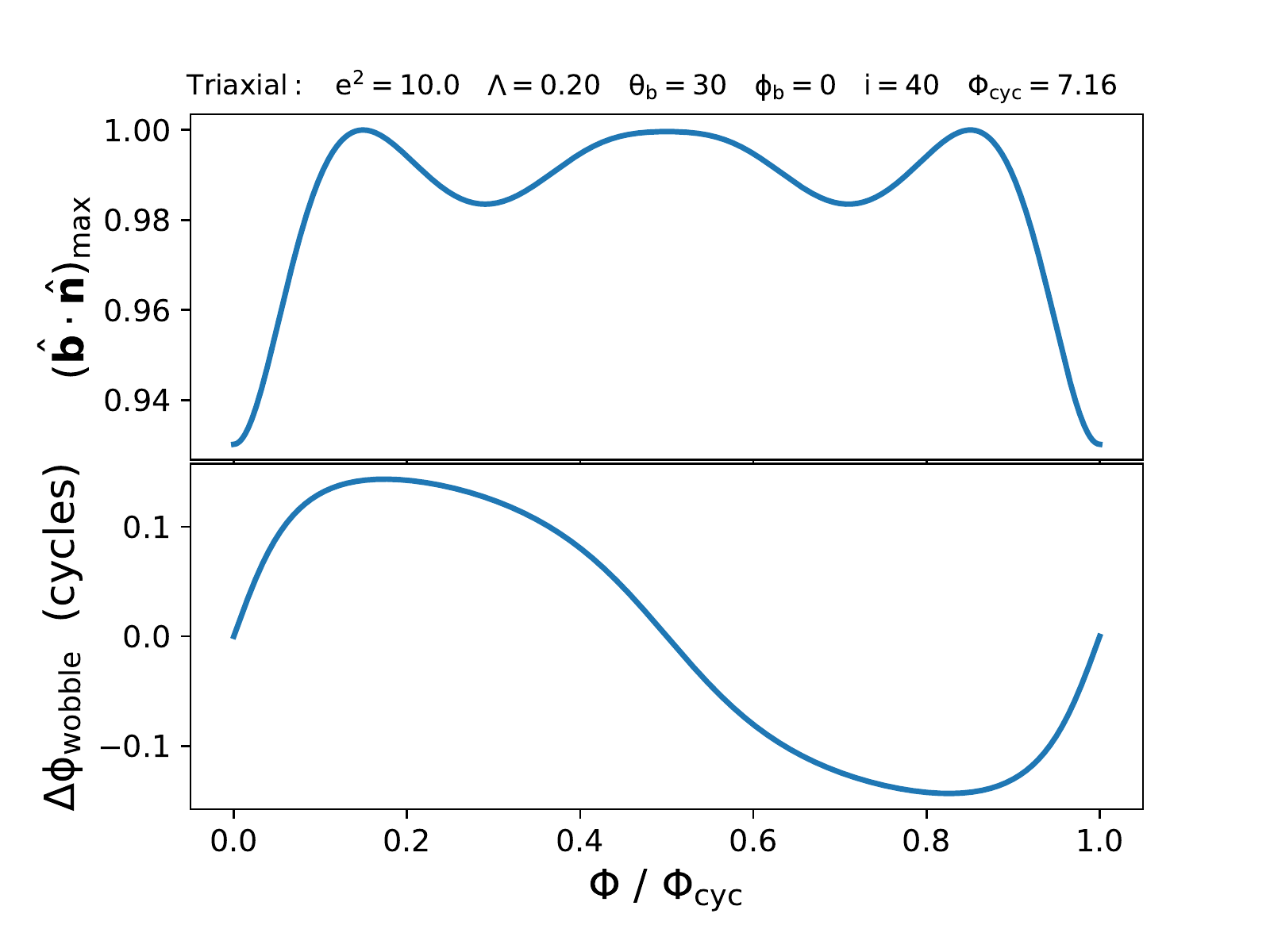}
  \includegraphics[width=\figwidthtwo]{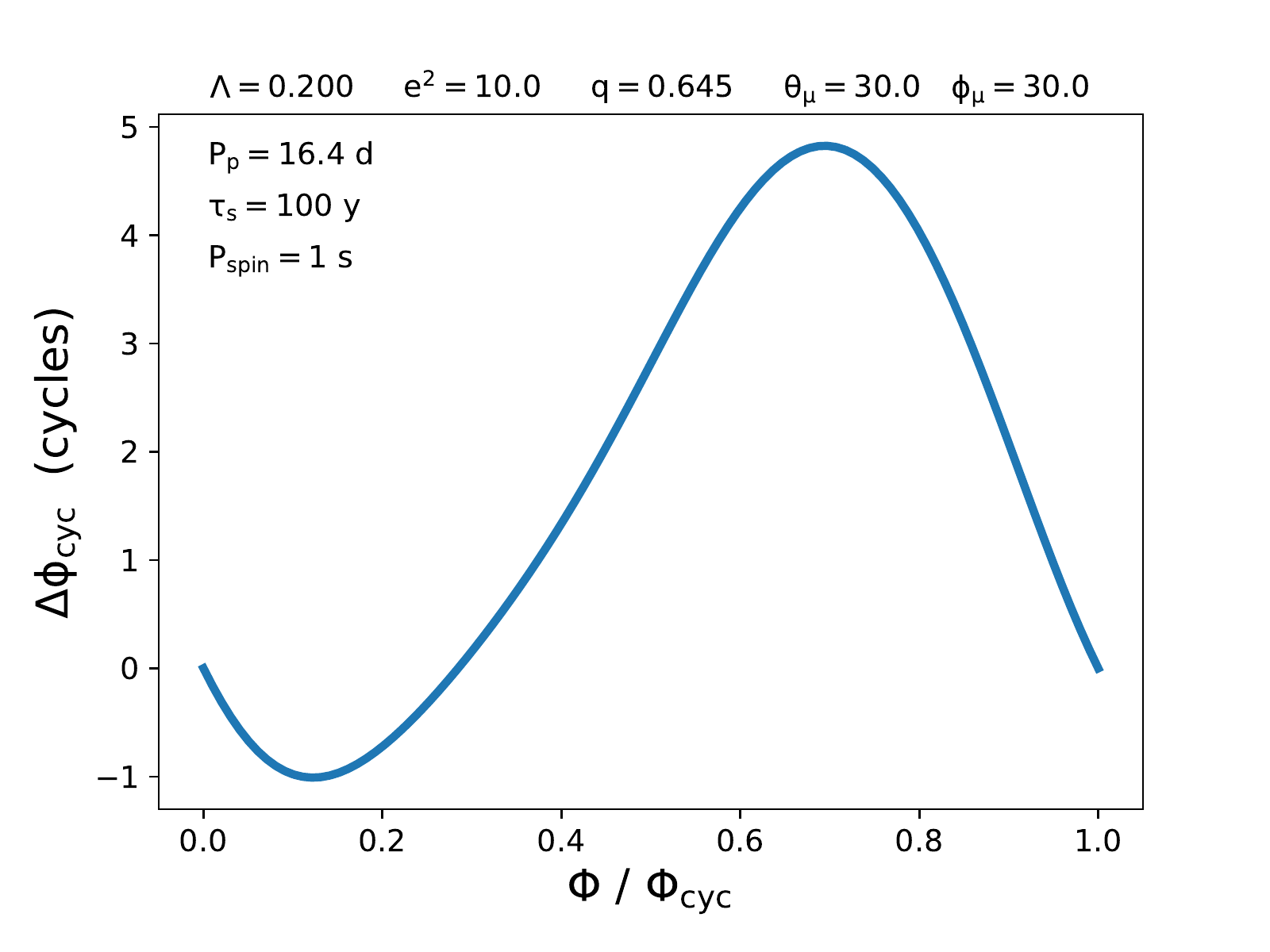}
\caption{Timing perturbations from triaxial precession with   $e^2 = 10$  and a precession amplitude $\Lambda = 0.2$~rad. 
Left:  Dot product $\bhat\cdot\nhat$ and spin phase perturbation due to beam wobble plotted over a precession cycle. 
Right:  Cyclical timing variation due to the magnetic torque calculated for a 16.4~d precession period and  a 100~yr spindown time.   A spin period $\Pspin = 1$~s is also used. 
}
\label{fig:toavariations}
\end{figure}

\subsection{Cyclical Timing Variation from the Magnetic Torque}
\label{sec:precession_timing_residuals}

The  steady spin down of a NS from the mean magnetic torque is well known but arrival times
will  vary cyclically from the variation in torque over a precession cycle \citep[e.g.][]{1993ASPC...36...43C, 2006MNRAS.365..653A}.   
The torque depends on the angle between  the magnetic moment $\muhat$ and the instantaneous spin axis  
$\lhatvec$ and so the  spin-rate derivative $\nudot$ that is a function of   $\muhat\cdot\lhatvec$ will show  both secular and cyclical variability.  

As in Paper~I, the magnetic  moment is oriented along the unit vector,
$\muhat=\ehat_1\muhatone+\ehat_2\muhattwo+\ehat_3\muhatthree$, 
in the rotating frame of reference, where the quantities  $\muhat_j \equiv \muhat\cdot \ehat_j$  are the direction cosines with respect to the principal axes,  $\ehat_j$  with $j=1, 2, 3$.   
Using Eq.~\ref{eq:lvhsol} and
 integrating $\nudot$ twice gives the spin phase  vs. time that includes secularly increasing terms  and the cyclical component of interest here, which is  expressed in cycles of spin phase instead of time,
$\Dphicyctorque = \nu \Dtcyctorque(\Phi)$,  as a function of precession phase, $\Phi$.    The full derivation is given in Paper~I (\S~2.4 and Appendix A) and  is  summarized in Appendix~\ref{app:cyc_torque} of this paper with slightly different notation. 

The result over time spans much shorter than the spindown time $\taus$ is
\be
\Dtcyctorque(\Phi)  
	&=&
	-  A_{\Dtcyctorque} \Lambda \cycfunc(\Phi, \Lambda, e^2, \muhat),
\label{eq:Dtieqn}
\ee
where   the characteristic amplitude is 
$A_{\Dtcyctorque} =  \Pp^2 / 8\pi^2\taus$ and $\cycfunc(\Phi, \Lambda, e^2, \muhat)$ gives the variation over
$\Phi$ (See Eqs.~\ref{appeq:toa_A} and \ref{appeq:toa_Q}).     For axisymmetric or small-amplitude precession,
the dependence on $\Phi$ (or, equivalently, time)  becomes sinusoidal in form as discussed in \S\ref{sec:unmodeled} and shown in Appendix~\ref{app:cyc_torque}.

The cyclical TOA variation is large for large precession periods and small spindown times, the latter applicable to young magnetars.    For the example of $\Pp = 16.4$~d and $\taus = 100$~y, 
$A_{\Dtcyctorque} \sim 8$~s, so even small values of $\Lambda$ can yield sizable phase perturbations that could mask a spin period $\sim$seconds.

Figure~\ref{fig:toavariations} (right panel) shows an example  timing variation (expressed as a phase
$\Dphicyctorque$) over a precession cycle corresponding to the beam-wobble case shown in the left panel. 
For other combinations of angles, the curves can be quite different.    

Figure~\ref{fig:torque_wobble_multiple} shows the torque angle $\cos^{-1}(\lhatvec\cdot\muhat)$, which determines the torque variation,  plotted against the phase variation from beam wobble  $\Dphiwobble$ 
for multiple values of the beam azimuthal angle $\phi_{\rm b}$. 
The cases shown are for a triaxial star with $e^2 = 10$ and $\Lambda = 0.2$   and polar angles
 $\theta_{\rm b} = \theta_{\mu} = 30^{\circ}$.   The pattern is asymmetric for $\phi_{\rm b} > 0$ and evolves slowly  until $\phi_{\rm b} \gtrsim 60^{\circ}$ for the polar angles.   The thickness of the curves corresponds to the amplitude of the beam precession modulation function. The thinnest parts correspond to the precession phases where bursts are much less likely to be detected.    The left panel is for a smaller Gaussian beam width of 3~deg (FWHM) and the right panel is 
 for a much larger 50~deg beam width.  The latter case, which shows a large amplitude throughout the precession cycle,  is inconsistent with observations of the slow periodicities in \Rone\ and \Rthree, which show distinct gaps where no bursts are detected.   For the 3~deg beam, the amplitude is large for only a small restricted part of the precession cycle.   
 
As already commented, magnetic field decay during the FRB-emitting phase of a young magnetar will reduce the amplitude of the cyclical phase variation, making burst periodicities more likely to be detected unless they are hidden by phase jitter, as discussed in \S~\ref{sec:decohere}.

\begin{figure}[t!] 
   \centering
   \includegraphics[width=\figwidthtwo]{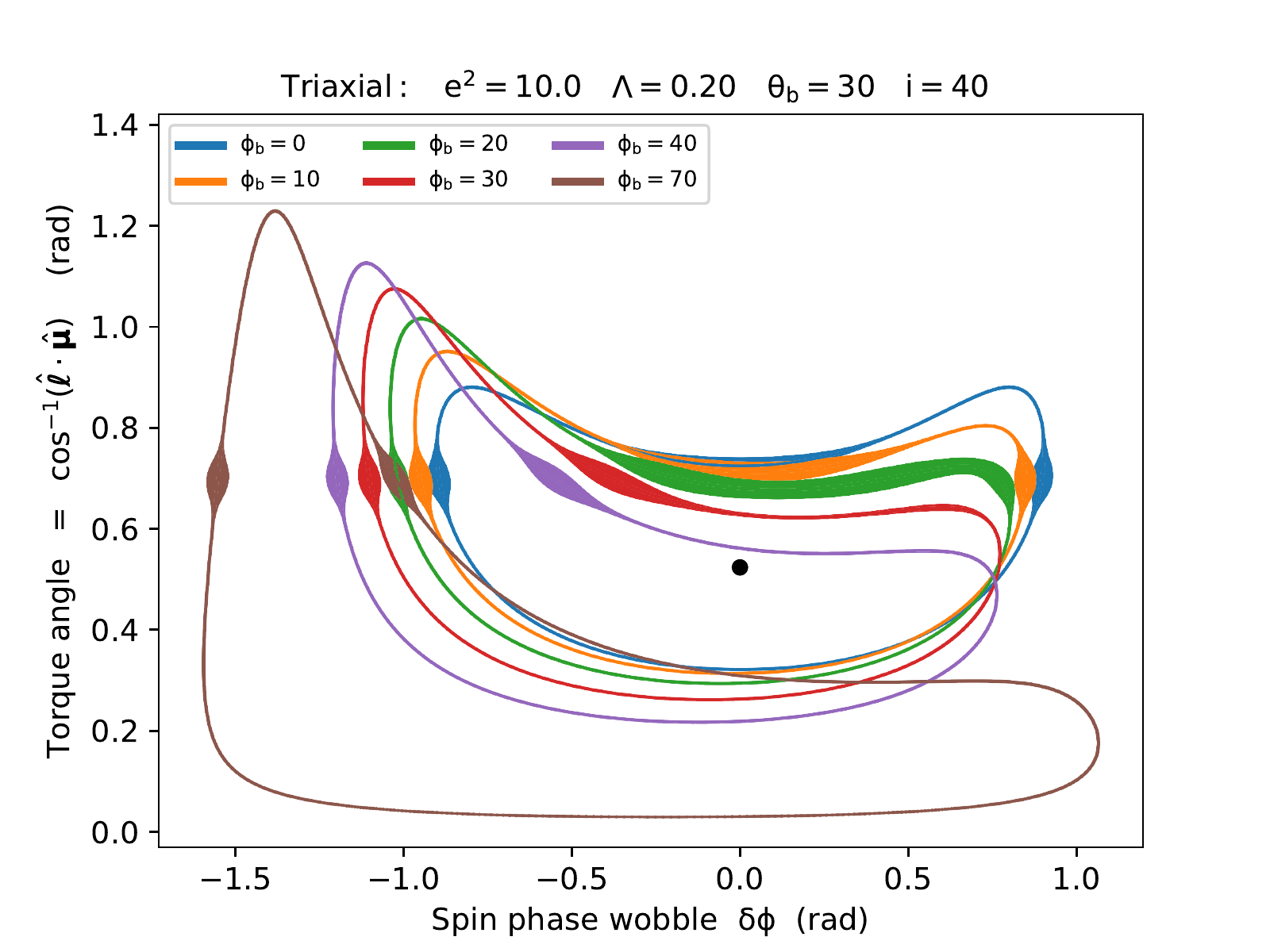}
   \includegraphics[width=\figwidthtwo]{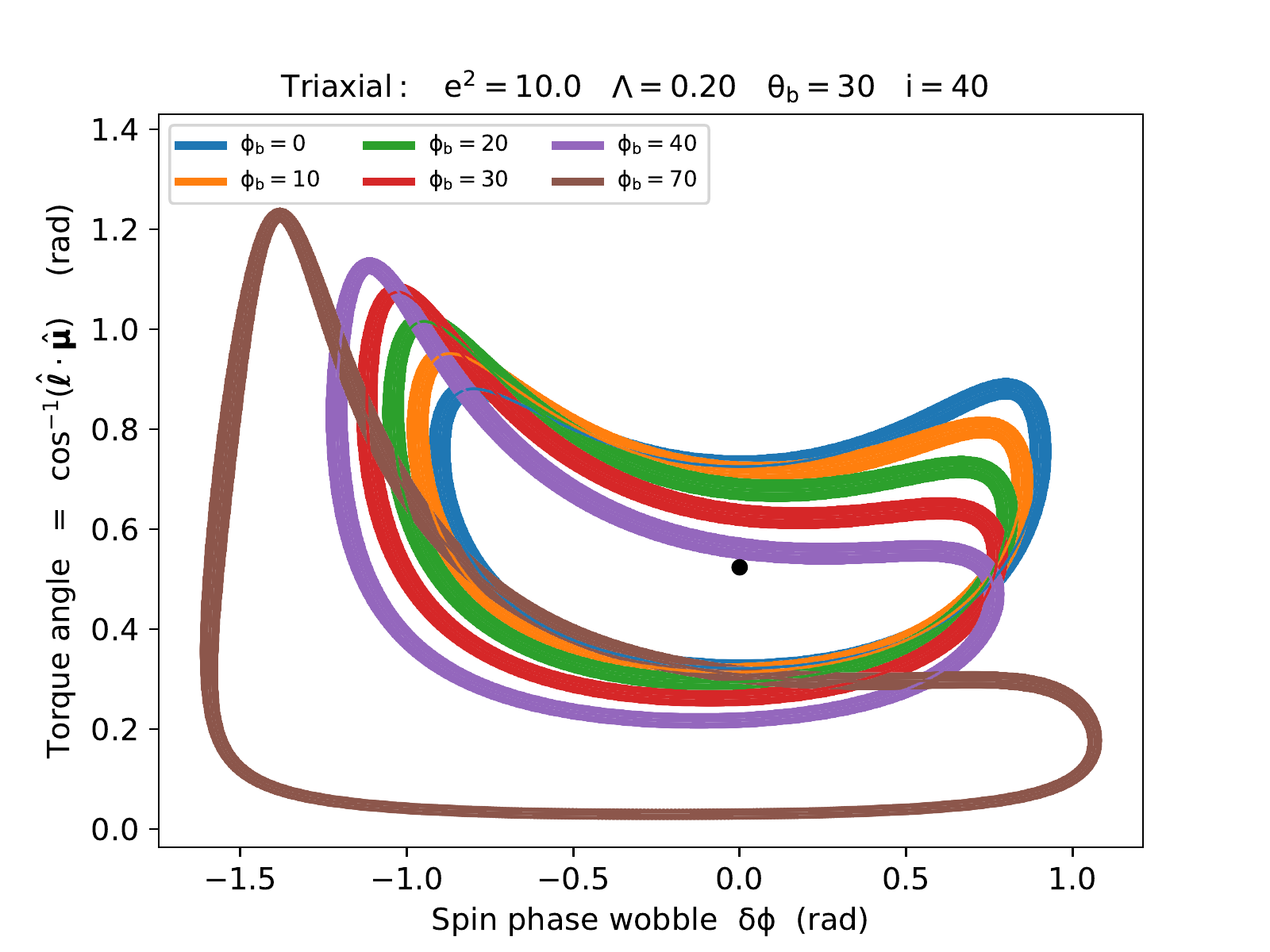}
     \caption{Torque angle $\cos^{-1}(\lhatvec\cdot\muhat)$ vs the wobble phase $\delta\phi$ for  
     $\Lambda = 0.2$, $e^2=10$, and $\theta_{\rm b} = 30^{\circ}$ for multiple values of the azimuthal beam angle, $\phi_{\rm b}$, as labeled.    The black circle indicates values for a non-precessing object 
     ($\Lambda = 0$) with azimuthal beam direction $\phi_{\rm b} = 0$.   The line widths scale with  the amplitude of the beam precession modulation function evaluated for a Gaussian beam function. 
     Left:  3~deg beam width.
     Right: 50~deg beam width.
     }
   \label{fig:torque_wobble_multiple}
\end{figure}

\clearpage
\section{Application of Triaxial Precession to Periodicity Detection}
\label{sec:applications}

We now consider periodicity detection of precessing objects using the detection statistic 
defined earlier (Eq.~\ref{eq:detstat}) applied to simulated data. 
To evaluate $\Dhat$ for different precession parameters, only  the peak burst amplitudes and residual phases are needed.  
The necessary data are generated as follows:
\begin{enumerate}
\itemsep -2pt
\item Choose the number of bursts $\Nb$ to Monte Carlo (MC) over a single precession cycle $\Phicyc$.  This is intended to be much smaller than the number of spin periods in the cycle, as is consistent with observations so far, because the emission is either zero (null) or weak in most periods. 
\item MC $\Nb$ values of precession phase, $\Phi_j, j = 1, \cdots, \Nb$.   
\item From $\Phi_j$ calculate the wobble and cyclic-torque phases ${\Dphiwobble}_j$ and ${\Dphicyctorque}_j$.
\item MC $\Nb$ amplitudes $a_j$ using a log-normal distribution with parameters that give 
unit mean and unit  modulation index,  $m_a = \sigma_a / <a> = 1$.  We allow there to be a specified  fraction $\fnull$ of `null' pulses like those seen for pulsars\footnote{While pulsar nulls are sustained for some number of contiguous pulses, we randomly turn bursts off independently because we are not interested in a more complex model for nulling that would require additional  parameters.}.
\end{enumerate}
With real data, we would measure arrival times  $t_j$ and  calculate phases $\phi_j$ using a 
model for the deterministic contributions to the phase, as in Eq.~\ref{eq:phimodtotal} and \ref{eq:phimodfit}.  Subtracting the fit yields phase residuals to use in the detection statistic with
weights given by the burst amplitudes for bursts that exceed a threshold.     

To illustrate,  we assess the effects of precession  on periodicity detection using  only a spin model for the deterministic phase and we assume the secular part of the spindown is known.  Arrival times are then  simply pulse numbers $n_j$  for the $j = 0, \ldots, \Nb-1$ bursts combined with phase perturbations, 
\be
t_j = (n_j  + \Dphiwobble + \Dphicyctorque) \Pspin
\label{eq:tj}
\ee
with pulse numbers  calculated as
\be
n_j = {\rm int} (\Phi_j \Pp / \Pspin).
\ee
For each  trial value for the spin period (designated by the caret)  $\Pspinhat$ we have  trial phases,
\be
\phi_j(\Pspinhat) = t_j / \Pspinhat,
\label{eq:phijhat}
\ee
and  phase residuals $\delta\phi_j$ are the fractional part of $\phi_j$. 
 A grid search over  $\Pspinhat$  yields a maximum  of $\Dhat(\Pspinhat)$ at  the correct period if other effects allow the periodicity to be manifested in the data. 

Using the above approach  we calculate trains of burst amplitudes along with the beam precession modulation 
function $\Bscr(\Phi)$.          The final burst amplitudes are the product of the log-normal amplitudes with the BPMF and the null-amplitude window.

Figures~\ref{fig:train1} and \ref{fig:train2} show burst trains over multiple precession cycles for one case with  90\% nulls and $\phi_{\rm b}=0$  and a second case with 80\% nulls and $\phi_{\rm b} =  30$~deg, respectively. The top panel in each frame shows the BPMF, the middle panel the random burst amplitudes, and the bottom panel shows the amplitudes with the BPMF and null-burst window applied.  For the first case with 1000 total bursts of which 90\% are nulls,  the precession periodicity is not obvious and there is an insufficient number of large amplitude bursts to identify it.   With a smaller null percentage of $10^4$ bursts in Figure~\ref{fig:train2}, the periodicity is easy to identify.    Also, the  unequal spacing of the three peaks in the BPMF, including one that is only about 5\% of the largest,    is discernible with the large number of strong bursts.    Currently, none of the repeating FRBs has provided sufficient bursts to test whether such triple peaks occur, including \Rone\ with the 
$1652$ bursts  detected with the FAST telescope \citep[][]{2021Natur.598...267L}.

\begin{figure}[t!] 
   \centering
   \includegraphics[width=0.8\linewidth]{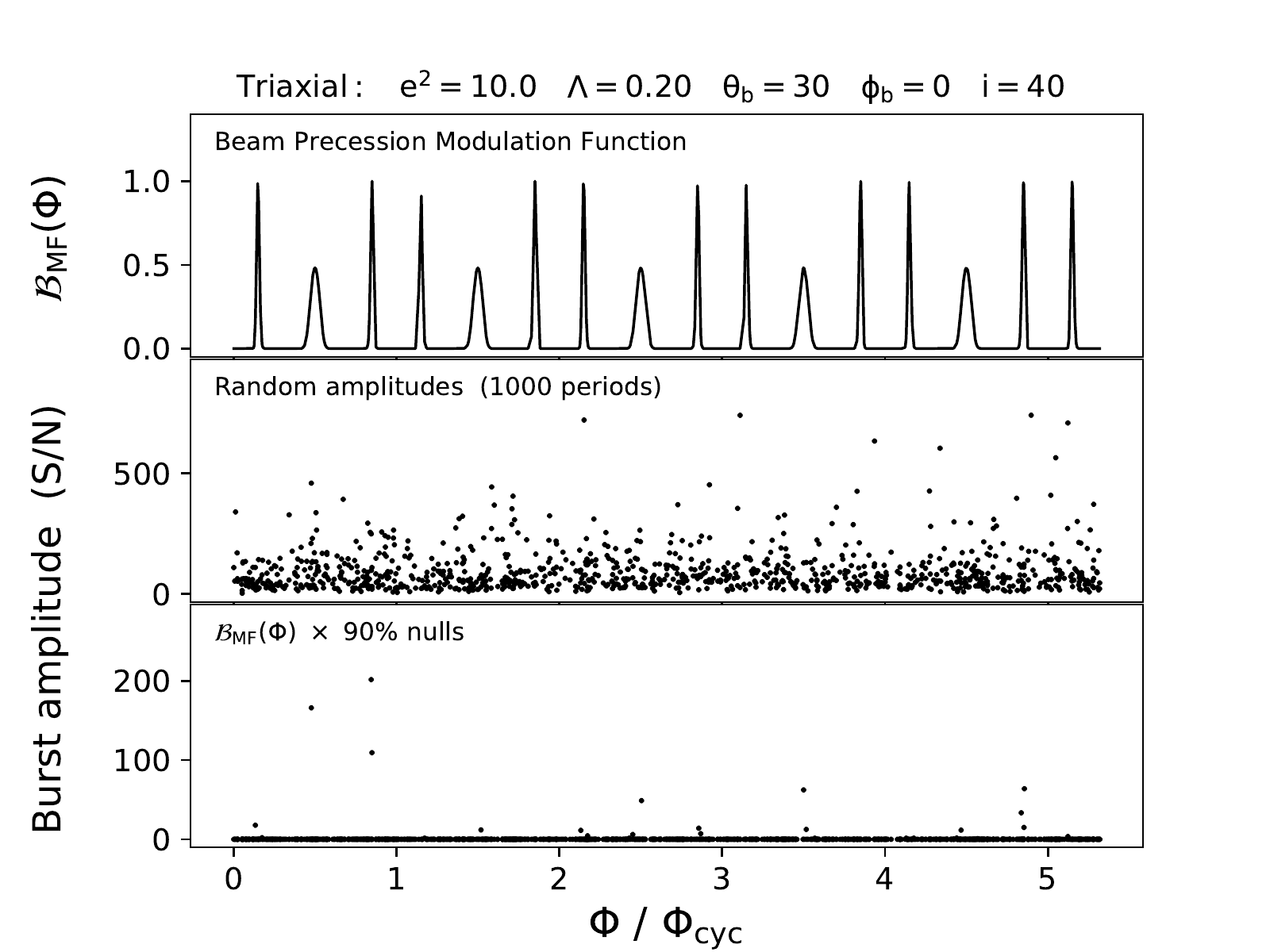}
     \caption{A burst sequence  with 90\% null pulses for $\phi_{\rm b} = 0$ with 5.3 precession cycles shown.
   Top:  Beam precession modulation function from wobble of the beam due to precession.
   It is evaluated using a Gaussian beam with $1/e$ width of 3~deg. 
   Middle:  Random burst amplitudes generated from a  skewed log-normal distribution 
   normalized to have unit mean and unity modulation index; these are then scaled by 
   the mean signal to noise ratio of ten. 
   Bottom:  Net burst sequence after applying the modulation function of the top panel and the indicated percentage  of nulls. In this case some of the bursts occur in a window that is not equally spaced with the window containing the brightest bursts.  }
   \label{fig:train1}
\end{figure}

\begin{figure}[t!] 
   \centering
    \includegraphics[width=0.8\linewidth]{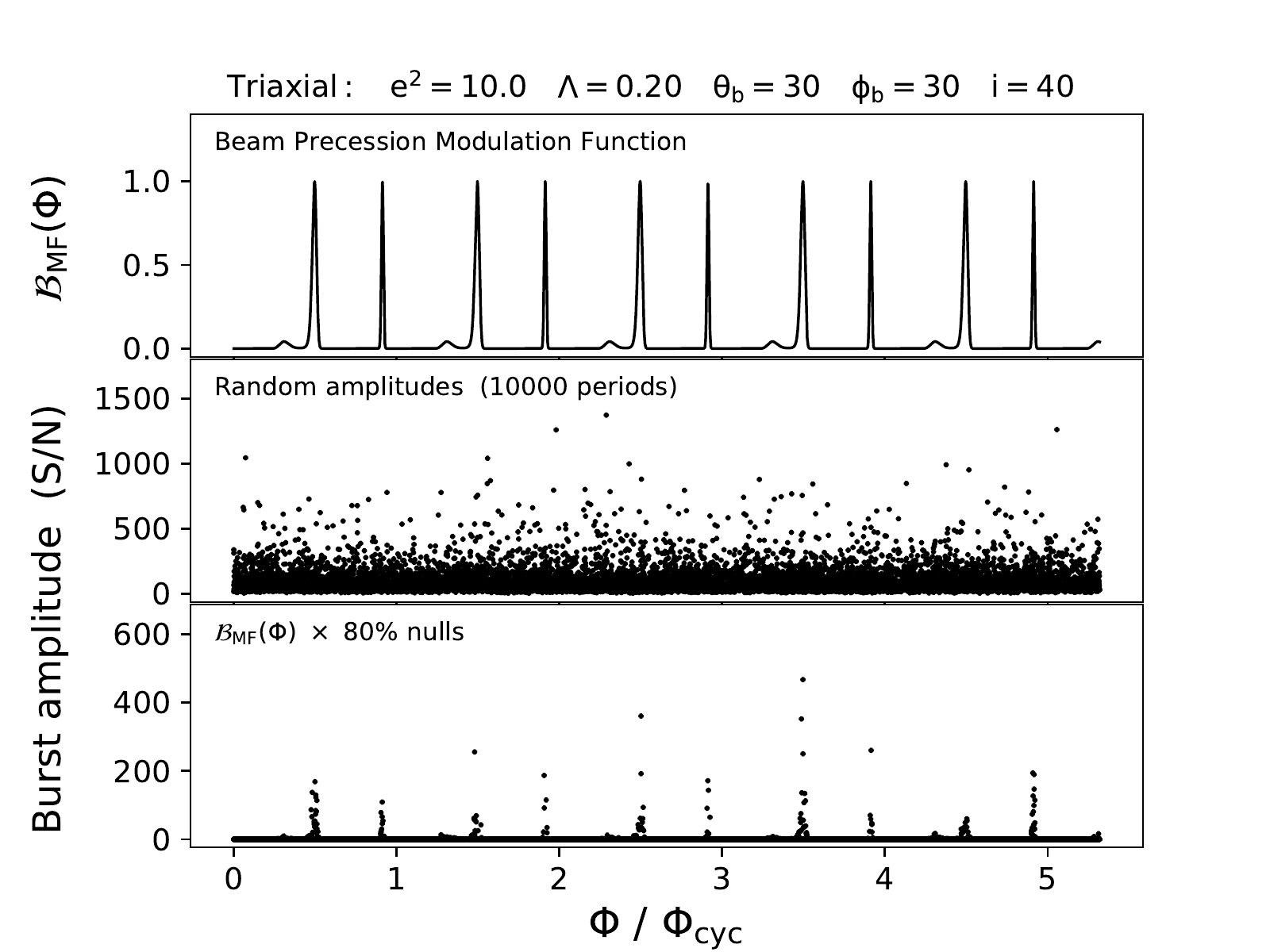}
     \caption{
     A burst sequence  with 80\% null pulses for $\phi_{\rm b} = 30$~deg. 
   Top:  Beam precession modulation function from wobble of the beam due to precession.
   It is evaluated using a Gaussian beam with $1/e$ width of 3~deg.
   Middle:  Random burst amplitudes generated from a  skewed log-normal distribution 
   normalized to have unit mean and unity modulation index; these are then scaled by 
   the mean signal to noise ratio of ten. 
   Bottom:  Net burst sequence after applying the modulation function of the top panel and the indicated percentage  of nulls.}
   \label{fig:train2}
\end{figure}

The detectability of the spin periodicity under different conditions is represented in
Figure~\ref{fig:D4}, which  shows the detection statistic for examples with and without precession and with and without phase jitter and contributions from the cyclical spindown torque.    The results are shown as a function of the trial spin period $\widehat{P}_{\rm s}$ used to evaluate the phase 
$\delta\phi = \Dphiwobble + \Dphicyctorque$
that is used in Eq.~\ref{eq:detstat}.   

The top left-hand panel shows cases with negligible  precession amplitude  for 1000 bursts spread over multiple precession cycles $N_{\rm pre} = T_{\rm pre} / P_{\rm p}$ where $T_{\rm pre}$ is the length of the time series.   The detection statistic maximizes at the true period where 
$\Delta \widehat{P}_{\rm s} = \widehat{P}_{\rm s} - {P}_{\rm s} = 0 $ and shows a main lobe that narrows as the bursts are spread over more precession cycles.   Even with amplitude variations of bursts (using a log-normal PDF with unit mean and modulation index of unity), the detection statistic reaches the maximum possible value (unity) because all bursts are strictly periodic. 

The bottom left panel shows  the results for axisymmetric precession with $\Lambda = 0.2$ for different 
combinations of number of bursts and spindown times, as indicated in the legend.   There is no difference in 
$\Dhat$ for $\Nb = 100$ or 1000 bursts if the cyclical torque is negligible, as it is for long spindown times.  
However, short spindown times correspond to many cycles of 
$\Dphicyctorque$
over the set of bursts that quench $\Dhat$. 

The right hand panels show cases with trixial precession with $e^2 = 10$ and $\Lambda = 0.2$.   
The top right panel shows $\Dhat$ vs inclination angle of the line of sight.    All cases show several maxima
in $\Dhat(\widehat{P}_{\rm s})$, including one at $\Delta \widehat{P}_{\rm s} = 0$ but also two others at offset
periods.    These result from the nutation of the beam, which causes bursts to be seen at multiple spin phases,
as shown in Figure~\ref{fig:torque_wobble_multiple}, that lead to multiple peaks in $\Dhat$. 
The bottom right panel shows how $\Dhat$ is quenched by phase jitter $\sigma_{\rm J}$ or by the spindown torque. 

\begin{figure}[t!] 
   \centering
   \includegraphics[width=0.8\linewidth]{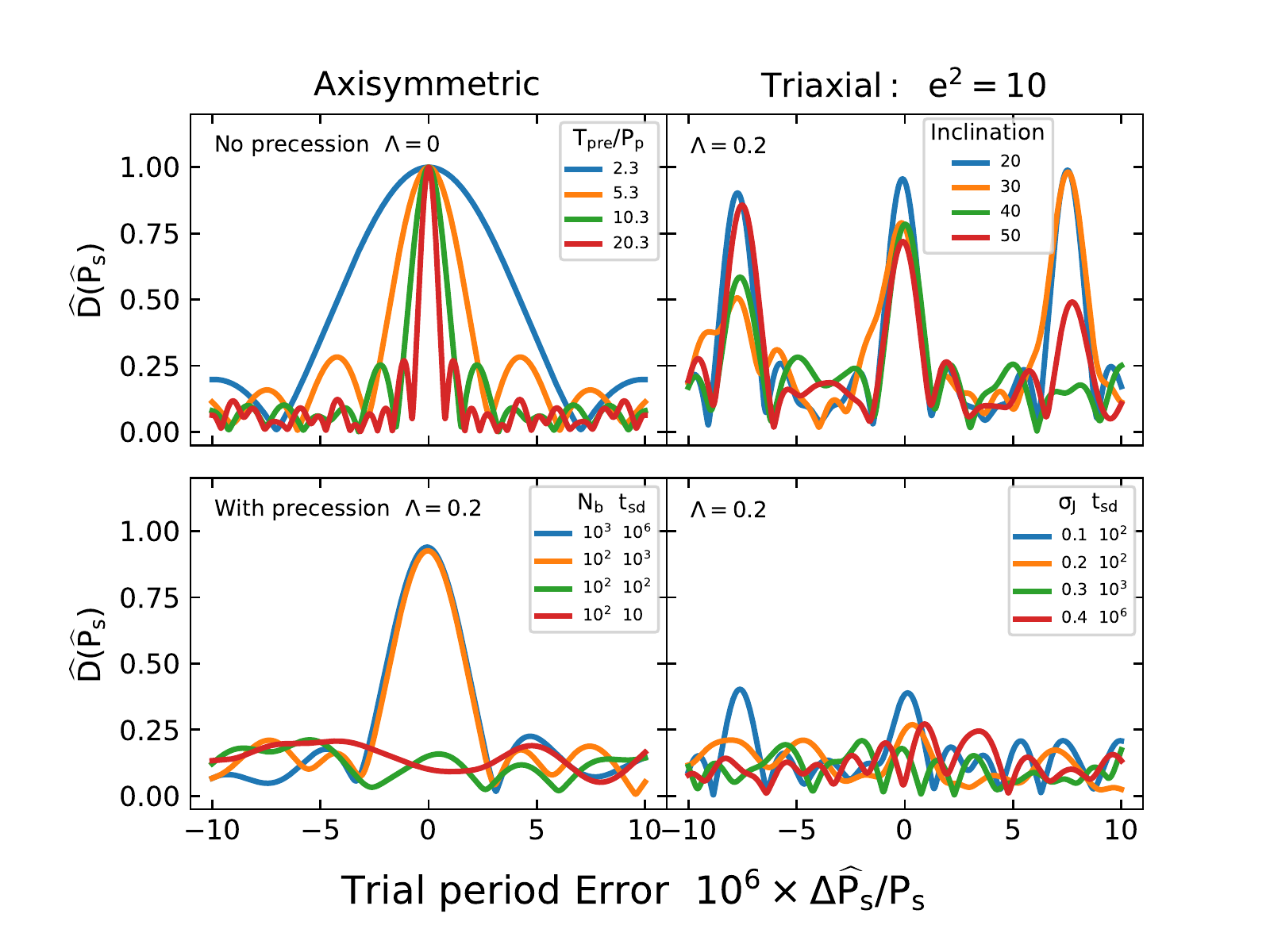}
   \caption{Detection statistic $\Dhat(\widehat{P}_s)$ vs $\Delta\widehat{P}_s /P_s$  for different precession cases for a Gaussian beam with $1/e$ width of 0.03 rad.  Details are given in the text.
   Top left: axisymmetric cases with negligible precession for 1000 bursts spread over different numbers of  precession cycles specified as $T_{\rm pre} / P_{\rm p}$ where $T_{\rm pre}$ is the time span and $P_{\rm p}$ is the precession period.  
   Bottom left: axisymmetric cases with $\Lambda = 0.2$ for different numbers of bursts spread over 
   5.3 precession cycles and for different spindown times (in years). 
   Top right:  Triaxial precession for different inclination angles and 1000 bursts spread over 5.3 precession cycles. 
   Bottom right: Triaxial precession for  5.3 precession cycles with 1000 bursts having different amounts of phase jitter $\sigma_{\rm J}$ and different spindown times (leading to different cyclical torque variations). 
   }
   \label{fig:D4}
\end{figure}

\section{Summary and Conclusions}
\label{sec:conclusions}

We have analyzed how a precessing FRB source of beamed radiation might show the slow periodicities seen in \Rone\ and \Rthree\  ($\sim 16$ and 160~days, respectively)  without showing any evidence for an underlying, faster spin periodicity. 

In burst sequences spread over multiple precessional cycles without any strong clustering over short time spans (e.g. hours),  several effects can make it difficult to see the fast periodicity.  These include timing variations from precession, either through wobble of the emission beam or from the variation in torque arising from dependence of the torque on the changing angle between the magnetic moment and the spin axis.   Noise in the spin rate from star quakes or stochastic changes in magnetic torque will also be important for this situation.  

Also included in our analysis is  phase jitter associated with the beamed radiation itself.  Phase jitter is needed to account for the absence of a fast periodicity when  a large number of bursts (tens to hundreds)  occur (and are detected) over a few hours and with spacings as short as a few seconds.    

Jitter could result from changes in direction of the beamed radiation related to multiple independently emitting beams.    However the required phase jitter $\gtrsim 0.3$~cycle, corresponding to more than 100~deg, is too large to be consistent with precession of the beam as a cause for the slow periodicities.   Beams this wide would allow bursts to be seen throughout the precession cycle, in contrast to the existence of gaps between the intervals when bursts occur.  

A natural explanation is that jitter is related to  a wide range of emission altitudes for different bursts.  Combined with strong relativistic beaming,   differences $\Delta r$ in altitude over a large fraction of the light-cylinder radius, $\rlc = c\Pspin / 2\pi$, correspond to phase variations 
$\Delta \phi \simeq 1 / \pi\sin\chi$  as a combination of  retardation and aberration,
where $\chi$ is the angle between the spin axis and observer's direction (which can vary over a precession cycle).    This amount of altitude variation can account for the absence of the fast periodicity in periodograms or power spectra of burst sequences.

Observed bursts from repeating FRBs have highly variable fluences and there are some activity windows (which correspond to favorable intervals of the precession cycle) when no or many fewer bursts are seen from 
\Rone\ and \Rthree.   These deficits might involve fading due to extrinsic scintillation or plasma-lensing or they could be caused by changes in the coherence of the radiation from processes that are independent of precession, such as a time-variable surface temperature of the NS that affects particle numbers and energies.

The slow periodicities observed so far also place constraints on precession.   Emission beams less than about 10 to 20~deg can be completely misdirected from the observer's direction for precession angles larger than the beam width, accounting for the absence of bursts in periodic data spans.      Observed bursts occur in slowly periodic phase windows that  are fairly large fractions $\simeq 0.25$ to 0.55 of a precession cycle, so they are not particularly constraining on the triaxiality of the stars or on relevant orientation angles of the angular momentum, beam, and line of sight.   Different combinations of angles combined with different degrees of triaxility ($e^2$) can yield single, double, triple, or quadruple  peaks in the beam precession modulation function (c.f. Figure~\ref{fig:bpmf1}) even when the beam is a unimodal, Gaussian like function.    The separations of these peaks may correspond to the apparent period of the observed slow periodicities, implying that the true precession period is larger.    

Eventually, data sets with much larger numbers of bursts may distinguish between these possibilities. 
In particular, bursts spanning many precession cycles can be folded (i.e. synchronously averaged) with the precession period.    The resulting shape will correspond to  that of the beam precession modulation function (Figure~\ref{fig:bpmf1}).  In addition, any radio-frequency dependence of the emission beam will be manifested in the BPMF shape.  While we do not know how the beam might vary with frequency, it is possible that some frequencies will be better than others for detecting large numbers of bursts.
However, it is also possible that the precession properties may vary, perhaps suddenly, when the figure of the star changes discontinuously as a result of magnetic driven shearing events.    

Finally, as noted in  Paper I,    the episodic aspects of FRBs that include a slow periodicity in two objects without any spin-related fast periodicity may evolve dramatically as a young magnetar ages, spins down, and some magnetic field components decay.   Such objects may emerge from an `FRB phase' into a phase similar to that of Galactic magnetars, which show episodic  radio emission that is periodic and very much like isolated, spin-driven pulsars. 

\vspace {0.2in}
The authors thank the referee for comments and suggestions that improved the presentation in the paper. 
SC and JMC acknowledge support from the National Science Foundation 
(NSF AAG-1815242) and are members of the NANOGrav Physics Frontiers Center, which is supported by the NSF award PHY-2020265.

\begin{appendix}
\section{Spindown  and  Spin Noise Estimates for Young Magnetars}
\label{app:spinnoise}

Pulsar spin rates decline  smoothly  from the average magnetic torque but also vary stochastically. 
The spin rate derivative is 
\be
\nudot = \frac{d\nu}{dt} = - \frac{k(2\pi)^2\mu^2 \nu^3 [1 - a(\muhat\cdot\lhatvec)^2]}{I c^3},
\label{appeq:nudot}
\ee
where $I = I_3$ is the principal moment of inertia.  As in Paper~I we use  values for the dimensionless constants $a=1/2$ and $k=2$ corresponding to the  spindown rate for a force-free magnetosphere,  $\nudot \propto 1 + \sin^2\theta$ with $\cos\theta = \muhat \cdot \lhatvec$.   This differs from the $\sin^2\theta$ dependence for  a vacuum magnetosphere, which is inconsistent with the torque on an aligned rotator \cite[][]{2012apj...746...60l}.
Expressing the magnetic moment $\mu =  B_{\rm d} R^3$ in terms of the surface dipole field and the stellar radius 
$R$ and using a moment of inertia $I = 10^{45}~{\rm g cm^2}\, I_{45}$
and a  fiducial dipole magnetic field strength $B_{\rm d} = B_{\rm d_{15}} 10^{15}$~G
we obtain 
\be
\nudot_{-15}  
= -\frac{\nudot } {10^{-15}}
\simeq 10^{6}\,{\rm Hz~s^{-1}}\, B_{\rm d_{15}}^2 \nu^3 .
\ee

Departures from smooth spindown include `glitches'  involving rapid increases in the spin rate $\Delta\nu / \nu$  accompanied by small changes in $\dot\nu$.   Pulsars also show stochastic spin noise manifested as much smaller changes in $\nu$ and $\dot\nu$ (of both signs  in some objects) and as  a red-noise process with a steep power spectrum in others.   Some pulsars show   discontinuous changes in the magnetospheric torque between two preferred states that last for weeks to months \citep[e.g.][]{2010Sci...329..408L}.   Here we consider the extension of pulsar spin noise to young magnetars as a means for estimating the minimum level of fluctuations.  Glitches will only exacerbate spin fluctuations that can inhibit the identification of the spin periodicity in sequences of FRBs. 

We  extrapolate  spin noise  to young magnetars using a scaling law for the rms residuals 
based on Galactic pulsars and magnetars
\citep[][hereafter SC10]{2010ApJ...725.1607S}; 
recent work \citep[][]{2017ApJ...834...35L,2019MNRAS.489.3810P,2020MNRAS.494..228L} has gotten similar results.  
This extrapolation requires the strong caveat that we simply do not know if the scaling law extends to younger, more rapidly rotating magnetars with larger magnetic fields, but it demonstrates that spin noise is likely to be important for the analysis of repeating FRBs if young magnetars are involved. 

The rms timing variation $\sigt$ scales as
$\sigt = C_{\rm spin} \, \nu^{\alpha} \, \vert \nudot_{-15}\vert ^{\beta} \, T^{\gamma}$ (SC10).
For  a data span length $T$ in years, spin frequency $\nu$  in Hz, and frequency derivative $\dot\nu = 10^{-15}\, {\rm Hz~s^{-1}}\, \dot\nu_{-15}$,  the coefficient is $C_{\rm spin} = 20 \Cfit$, where
$\Cfit = 11^{+7.2}_{-4.3}\, \mu s$ results from a second-order polynomial fit to timing data that accounts for quadratic spindown (from the  `CP+MAG'- fit in SC10's Table~1)  and the factor of 20 
corrects for the removal of spin noise by the fit; it is  based on simulations
\cite[][Table~2]{1980ApJ...237..216C} and applies to a mixture of random walks in 
$\nu$ and $\dot\nu$.  
Other parameters are 
$\alpha = -1.4\pm 0.2$,
$\beta = 1.13 \pm 0.07$ and
$\gamma=1.7\pm0.2$. 
 The large scatter about this relationship for different  objects  is characterized by a log-normal distribution with $\delta \equiv \sigma_{\ln \sigt / 20}(T) = 1.7\pm0.2$.
 The corresponding power spectrum $\propto f^{-x}$ with $x = 2\gamma+1 \simeq 4.4\pm 0.4$.
 

Using $\nudot_{-15}$ defined above,  the  rms timing variation is
\be
\sigt &=& 
10^{6\beta} 
C_{\rm spin}\,  
B_{\rm d_{15}}^{2\beta} \,
\nu^{\alpha+3\beta}\, 
T^{\gamma}
\simeq
 1326^{+870}_{-519}  \, s  \times \nu^2 \, B_{\rm d_{15}}^{2.3}\, T^{1.7} ,
\label{eq:sigtapp}
\ee
where $T$ is again in years and  for simplicity we have propagated the error on $\Cfit$ but not on the exponents. 
This approach, along with the assumption of  magnetic dipole radiation,
suffices for our goal of getting a qualitative assessment of the role of spin noise in periodicity detection. 

\section{Power Spectrum of Bursts with Phase Jitter and Nulling}
\label{app:spectrum}

\providecommand{\amean}{\langle a \rangle}
\providecommand{\abmean}{\langle a_b \rangle}

\renewcommand{\Nb}{N}

The spectrum in Eq.~\ref{eq:spec} of the burst sequence in Eq.~\ref{eq:sigmod1} is derived here for the simplest case where the periodic phase of an individual burst is modified by phase jitter.   Phase is measured in cycles.   
The time series of length $\Nb$ cycles is then, 
\be
I(\phi) =  \sum_{j=0}^{\Nb-1}   a_{j} A(\phi - j - \phi_j).
\label{appeq:sigmod1}
\ee
As described in the main text,  the  shape for an individual burst $A(\phi)$ is assumed identical for all bursts.
The  stochastic amplitudes   are statistically independent between bursts with mean and variance given by
\be
\langle a_{j} \rangle = \amean
\quad \text{and} \quad  
\langle a_{j} a_{j^{\prime}} \rangle = (1 + \ma^2) \langle a \rangle^2 \delta_{jj^{\prime}} ,
\ee
where $\delta_{jj^{\prime}}$ is the Kronecker delta,  $\ma = \sigma_a / \amean$ is the modulation index
 (rms amplitude divided by the mean), and angular brackets denote ensemble average. 
 To incorporate null bursts, those with $a_j = 0$, we adopt a probability density function (PDF)
 using a burst fraction $\fb$, 
\be
f_a(a) = (1- \fb) \delta(a) + \fb g_b(a),
\ee
where $\delta(a)$ is the Dirac delta function and non-null amplitudes follow a PDF $g_b(a)$  with mean  amplitude $\abmean$ and modulation index $\mb \sim 1$.   We then have
\be
\amean = \fb \abmean 
\quad \text{and} \quad
1 + \ma^2  =  (1 + \mb^2) / \fb ,
\ee
A small burst fraction $\fb\ll 1$ significantly increases the net modulation index $\ma$. 
Phase jitter is also assumed to be statistically independent between bursts with zero mean and
variance $\rmsj^2$.

The  Fourier transform (FT) of the burst shape is $\FTA$ for a Fourier kernel 
$e^{-2\pi i f \phi}$ (with $f$ in cycles per unit  phase).
Combined with the assumed statistical properties of the burst amplitudes
and phases, the spectrum is the squared magnitude of the FT of $I(\phi)$.  Using the Fourier shift theorem
on Eq.~\ref{appeq:sigmod1} we obtain
\be
S(f)  =  
	\langle\vert \widetilde I(f) \vert^2   \rangle 
	&= &
 	\vert \FTA \vert^2 
	\sum_{j=0}^{\Nb-1} \sum_{\jp=0}^{\Nb-1}
	\langle a_j a_{\jp} \rangle
	 e^{-2\pi i  f(j-\jp)}
	\langle e^{-2\pi i f(\phi_j - \phi_\jp)} \rangle
	\nonumber \\
	%
	%
	&=&
 	\amean^2 \Nb \vert \FTA \vert^2 
	\,
	\Bigl[ 
		\ma^2
		+
		\Nb^{-1}
		\Bigl\vert
			\sum_{j=0}^{\Nb-1} 
			 e^{-2\pi i  f(j-\jp)}
			\langle e^{-2\pi i f\phi_j} \rangle
		\Bigr\vert^2
	\Bigr] .
\label{appeq:spec1}	
\ee
When $\Nb \gg 1$, the summation in the last equality  averages to zero except at  harmonics  at and near 
(within $1/\Nb$) integer frequencies,  $f = \ell$. The squared sum is therefore a sum of such harmonics, each having a shape given by the squared `digital' sinc function [distinct from the continuous sinc function, $(\sin\pi x) / \pi x$], 
\be
\Delta_{\Nb}(f) =  \left[\frac{\sin (\pi \Nb f)}{\Nb\sin (\pi f)}\right]^2, 
\label{eq:appsinc2}
\ee
which is normalized to unit amplitude $\Delta_{\rm b}(0)  = 1$ and has a
width $\Delta f \simeq \Nb^{-1} \ll 1$ for large $\Nb$.     

The average $\langle \exp(-2\pi i f \phij)\rangle$ in  Eq.~\ref{appeq:spec1} is the characteristic function of $\phi_j$ that we term the jitter `form factor' and we evaluate for phase jitter having  a zero-mean, Gaussian PDF,
\be
\Jitt(f) 
=   \left\langle e^{-2\pi i f \phij)} \right\rangle 
= e^{-2(\pi f \rmsj)^2} .
\label{appeq:Jitt}
\ee
Alternative jitter distributions, including those with multiple modes, are easy to incorporate using their characteristic functions.  Including the effects of null pulses, the spectrum is
\be
S(f) = 
\fb (1 +  \mb^2 -\fb) \abmean^2  \Nb  \vert \FTA \vert^2   
\Bigl \{ 1 +  \Bigl[\frac{\fb \Nb \Jitt^2(f)} {1 +  \mb^2 -\fb} \Bigr] \sum_{\ell = 0}^{\infty} \Delta_{\Nb}(f-\ell) \Bigr \} .
\label{appeq:spec3}
\ee
The spectrum therefore includes a continuum term  superposed with  spectral lines, all of which are shaped by   the envelope function
$\vert \FTA \vert^2$ determined by the burst shape.  As the burst fraction $\fb$ decreases,  the spectral lines diminish relative to the  continuum term.    Large phase jitter implies $\Jitt(f) \to 0$, reducing spectral lines exponentially (for Gaussian jitter).

\section{Arrival Time Variation from Cyclical Torque}
\label{app:cyc_torque}

Free precession induces a cyclical variation in torque that adds to the slowly changing magnetic torque acting on a neutron star.   It results from the changing angle between the unit vectors for the  magnetic moment $\muhat$ and the instantaneous spin axis $\lhatvec$  over a precession cycle. 

The total spin phase perturbation from precession $\Dphi$ is obtained by integrating the spin-rate derivative
in Eq.~\ref{appeq:nudot} taking into account the variation of $\lhat$ over a precession cycle; details are given  in \S2.4 and Appendix A of  Paper~I. 


The result is a general expression  for the precessional time of arrival (TOA)  variation, $\Delta t$ (Eq.~59 of Paper~I). Secular terms that are linear and quadratic in time (note spin phase  $\phi$ is used as a proxy for time in Paper~I)  add to the cyclical term of interest here.  Here we consider only  
oblate stars  with $0 < \Lambda\sqrt{1+e^2} < 1$;   another solution for prolate stars is also presented in Table~1 of Paper~I that gives qualitatively similar results.   The cyclical part of the TOA variation is 
\be
\Dtcyctorque(\Phi)  
	&=&
	- A_{\Dtcyctorque} \Lambda \cycfunc(\Phi, \Lambda, e^2, \muhat)
\label{appeq:Dteqn}
\ee
where the variation over precession phase is 
\be
\cycfunc(\Phi, \Lambda, e^2, \muhat) 
	&=&
	\Lambda
	\left[\muhattwo^2 (1+e^2) - \muhatone^2 - e^2\muhatthree^2 \right]   C_1(\Phi|q)
		+2 \muhatone \muhattwo \Lambda \sqrt{1+e^2}   C_2(\Phi|q) 
	\nonumber \\
	&& \quad + 
	2\muhatthree\sqrt{1-\Lambda^2}
	\left[ \muhatone C_3(\Phi|q)-\muhattwo \sqrt{1 + e^2} C_4(\Phi|q) \right]~,
\label{appeq:toa_Q}
\ee
(recall $q = \Lambda \sqrt{1+e^2}$ for oblate stars, as given in Table~1 of Paper~I) 
and the leading coefficient is 
\be
A_{\Dtcyctorque} 
= \frac{ak\mu^2 }{I c^3 \left(d\Phi / d\phi \right)^2} 
=  \frac{\Pp^2}{8\pi^2\taus}.
\label{appeq:toa_A}
\ee
The second form in Eq.~\ref{appeq:toa_A} involves  the precession period $\Pp$ and the spindown time $\taus$ for magnetic dipole radiation.

The functions  $C_i(\Phi \vert q)$ in Eq.~\ref{appeq:toa_Q} are derived in Appendix A of  Paper I.   For $q =0$, applicable to axisymmetric  precession,  they are sinusoidal in precession phase $\Phi$, 
\be
C_1(\Phi \vert 0)=\frac{\cos 2\Phi -1}{8} ,
\quad
C_2(\Phi \vert 0)=-\frac{\sin 2\Phi}{8} ,
\quad
C_3(\Phi \vert 0)=1-\cos\Phi ,
\quad
C_4(\Phi \vert 0)=\sin\Phi .
\label{appeq:Cizeroq}
\ee
The functions $C_1(\Phi|q)$ and $C_2(\Phi|q)$ have periods equal to half of the precession period while   $C_3(\Phi|q)$ and $C_4(\Phi|q)$ have periods equal to a full precession cycle.

As an example, for $\Pp = 16.4$~d and $\taus = 100$~y with $\Lambda = 0.2$, the cyclical term has an amplitude
$A_{\Dtcyctorque} \Lambda \sim 1.6$~s, a substantial variation that would need to be fitted for in order to find the spin periodicity in a data set longer than about $\Pp/4$ for spin periods of order a second or less.   For shorter data spans, the precession variation would be absorbed in the apparent spin period and period derivative of a fit to arrival times or in a Fourier analysis that includes an acceleration component \citep[e.g.][]{2002AJ....124.1788R}.

Eq.~\ref{appeq:Dteqn} is fully nonlinear in $\Phi$ and $\Lambda$, but simplifies considerably for small amplitude precession, $\Lambda\sqrt{1+e^2}\ll 1$, in which case 
the only terms that survive to linear order in $\Lambda$ are those involving $C_3(\Phi)$ and $C_4(\Phi)$. 
Using   Eqs.~\ref{appeq:Cizeroq} we obtain 
\be
 \Dtcyctorque(\Phi)
\approx
- \left(
 \frac{\Lambda \Pp^2 }{8\pi^2\taus} 
 \right)
\muhatthree \sqrt{\muhatone^2+\muhattwo^2(1+e^2)} \cos(\Phi-\Phi_0) ,
\label{appeq:low_amp_dt}
\ee  
where the phase $\Phi_0$ is given by
\be
\cos\Phi_0=\frac{\muhatone}{\sqrt{\muhatone^2+\muhattwo^2(1+e^2)}} , 
\quad\quad
\sin\Phi_0=\frac{\muhattwo\sqrt{1+e^2}}{\sqrt{\muhatone^2+\muhattwo^2(1+e^2)}} .
\ee
For  axisymmetric precession of an oblate star ($e^2 = 0)$,  Eq.~\ref{appeq:low_amp_dt} is consistent with the estimate given in \S~\ref{sec:unmodeled}.

\end{appendix}



\end{document}